\RequirePackage{fix-cm}
\documentclass[natbib]{svjour3} 
\smartqed 
\usepackage{graphicx}
\usepackage{amsmath}
\usepackage{amssymb}
\usepackage{gensymb} 
\usepackage[utf8]{inputenc} 
\usepackage{rotating} 
\usepackage{caption} 
\journalname{Celestial Mechanics and Dynamical Astronomy}
\begin{document}

\title{Long-term orbital dynamics of trans-Neptunian objects}

\author{Melaine Saillenfest$^1$}
\authorrunning{M. Saillenfest} 

\institute{
   $^1$ IMCCE, Observatoire de Paris, PSL Research University, CNRS, Sorbonne Universit{\'e}, LAL, Universit{\'e} de Lille, 75014 Paris, France\\
   \email{melaine.saillenfest@obspm.fr}
}

\date{\emph{This article is dedicated to Giovanni B. Valsecchi.}\\[0.5cm]
Received: 13 September 2019 / Accepted: 16 January 2020}

\maketitle

\begin{abstract}
   This article reviews the different mechanisms affecting the orbits of trans-Neptunian objects, ranging from internal perturbations (planetary scattering, mean-motion resonances, secular effects) to external perturbations (galactic tides, passing stars). We outline the theoretical tools that can be used to model and study them, focussing on analytical approaches. We eventually compare these mechanisms to the observed distinct populations of trans-Neptunian objects and conclude on how they participate to the sculpting of the whole distribution.
   
   \keywords{trans-Neptunian object \and orbital dynamics \and chaos \and resonance}
\end{abstract}

\section{Introduction}
\begin{figure}
      \centering
      \includegraphics[width=\textwidth]{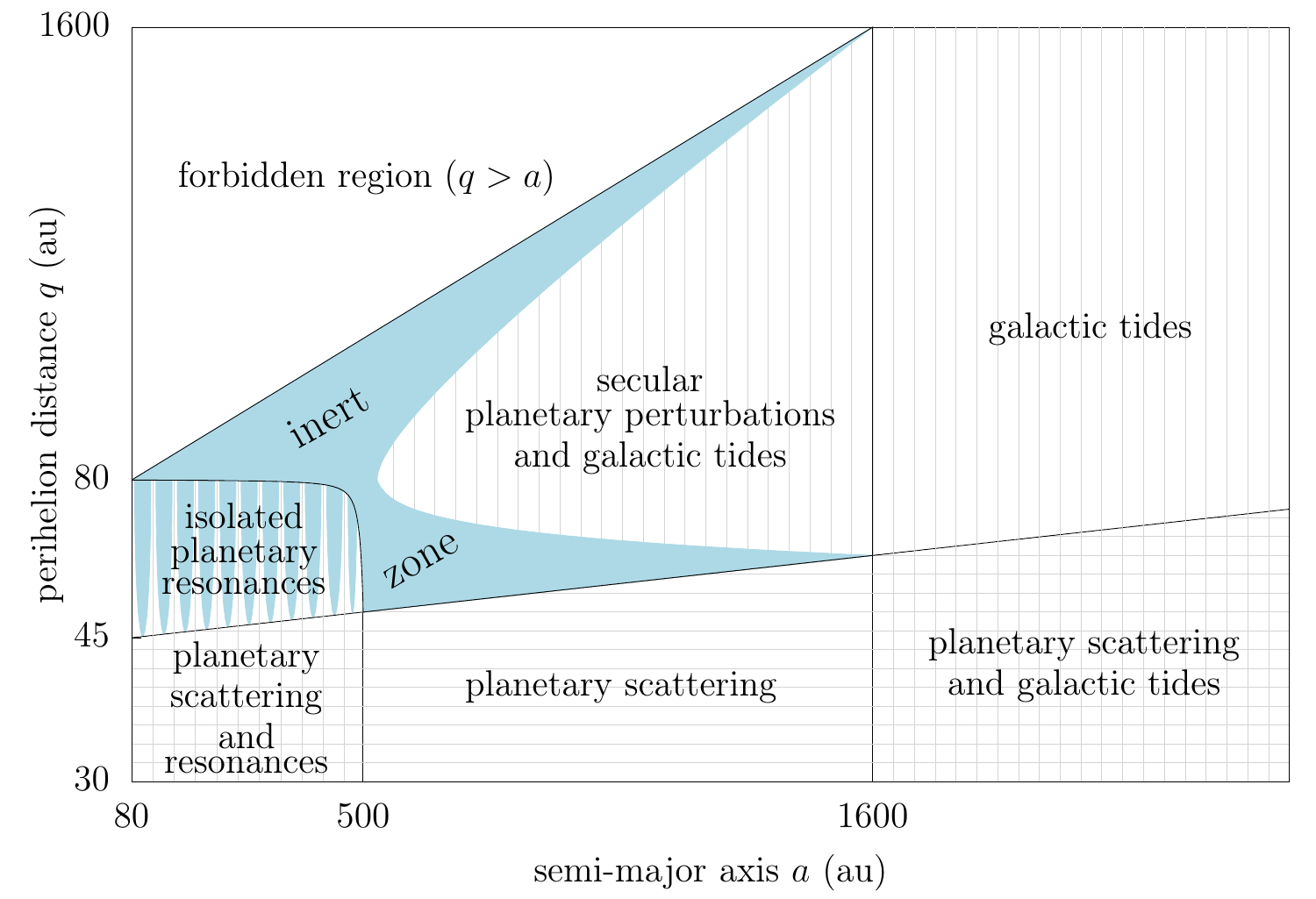}
      \caption{Schematic view of the regions where the different mechanisms of orbital dynamics are dominant, adapted from \cite{SAILLENFEST-etal_2019}. Hypothetical planets beyond Neptune are not taken into account in this picture. The planetary scattering process makes small bodies move horizontally, whereas the planetary mean-motion resonances, the planetary secular effects, and the galactic tides, make them move vertically. Passing stars produce sporadic jumps in the $(a,q)$ plane. The blue region is characterised by extremely long transport timescales: apart from precession, the objects that it contains hardly move at all during the solar system lifetime. The orbital inclination of small bodies is not represented here for simplicity, although the precise limit of the different regions depends on its value as well (see Sects.~\ref{sec:sec} to \ref{sec:inert}). As shown throughout this review article, variations of $q$ are almost systematically accompanied by inclination changes.}
      \label{fig:zones}
   \end{figure}

   From the prediction of their existence by \cite{EDGEWORTH_1949}, \cite{KUIPER_1951}, and \cite{OORT_1950}, and up to the most recent discoveries, the populations of objects beyond Neptune (the ``trans-Neptunian'' objects) never stopped showing how incredibly rich their orbital dynamics is. Their trajectories involve mechanisms as diverse as close encounters, chaotic scattering driven by resonance overlap, secular effects from the giant planets, isolated mean-motion resonances with Neptune, quasi-integrable cycles from the galactic tides, and even random impulses due to close passages of stars. Based on previous works (that we will specify later), one can get an idea of where these mechanisms are most efficient. A schematic picture of the different regions obtained is given in Fig.~\ref{fig:zones} in the plane of the semi-major axis and the perihelion distance.
   
   As shown in Fig.~\ref{fig:zones}, all dynamical regions are linked, allowing small bodies to switch between very distinct kinds of dynamics. For instance, a trans-Neptunian object initially evolving smoothly in an isolated mean-motion resonance can be transferred to the unstable scattering region, where its semi-major axis can grow enough for galactic tides to lift its perihelion distance, turning off the scattering; it can then be affected by a stellar passage, end in the inert zone, etc. For this reason, we think that a review article describing each dynamical mechanism involved in a unified picture would be useful for the community. We aim to make such an article available, and to provide the mathematical and numerical tools that can be used to study these mechanisms and their connections.
   
   The questions that we will address are mainly about the dynamics itself: what is the nature of the dynamics? Where is it produced in the space of orbital elements? Which orbital changes can it produce? How to model it analytically or numerically? We will avoid the complex task of comparing models to observations, and drawing conclusions about how to tweak the models in order to make them better represent the current state of the solar system. Yet, even though we will rather refer to the types of orbital dynamics, and not to the observed classes of trans-Neptunian objects (whose limits are sometimes fuzzy and not systematically linked to the dynamics), we will always try to draw a parallel between dynamics and observed objects. As such, we cannot avoid using terms like ``centaurs'', or ``Halley-type comets'', which may puzzle the readers non specialised in solar system dynamics. Therefore, we decided to begin this review by a short historical note about trans-Neptunian objects.
   
   This review article is organised as follows. In Sect.~\ref{sec:hist}, we recall the main historical landmarks of our knowledge of trans-Neptunian objects. The basic terminology is introduced. Then, Sect.~\ref{sec:adiff} is dedicated to the planetary scattering process. Sections~\ref{sec:sec} and \ref{sec:res} present the secular and resonant dynamics driven by the giant planets and introduce semi-analytical models used to determine their range of outcomes. Section~\ref{sec:gt} is dedicated to the perturbations from the galactic tides. Section~\ref{sec:inert} further investigate the dynamics in the intermediate regime between planetary-dominated and galactic-dominated dynamics, unveiling the dynamical structure of the limit between the Kuiper belt and the Oort cloud. Section~\ref{sec:stars} is dedicated to passing stars. Finally, Sect.~\ref{sec:sculpt} summarises how all these mechanisms participate to the sculpting of the observed populations of trans-Neptunian objects in the context of our current understanding of the formation of the solar system.
   
\section{Historical perspective}\label{sec:hist}
   The question of the existence of small solar-system bodies beyond Neptune is closely related to the origin of comets. Comets are traditionally classified into short-period comets (period $P<200$~yrs) and long-period comets ($P>200$~yrs). As recalled by \cite{WEISSMAN_1995}, this distinction is mostly historical: it roughly corresponds to the maximum time in the past up to which periodic comets can be identified using archive data. It rapidly appeared, however, that the differences between the orbits of short- and long-period comets do not limit to their periods. Indeed, the orbits of long-period comets are distributed almost isotropically in space, whereas the orbits of short-period comets are much more packed near the ecliptic plane. Prompted by another clear dichotomy in the distribution of orbital inclinations, astronomers further divided short-period comets into Jupiter-family comets ($P<20$~yrs) and Halley-type comets ($P>20$~yrs). Indeed, Jupiter-family comets, whose dynamics strongly depend on their interactions with Jupiter, have very small orbital inclinations, whereas the inclinations of Halley-type comets follow a broader distribution that even extends to retrograde orbits \citep{LEVISON_1996}. Such differences between long-period, Halley-type, and Jupiter-family comets, were immediately understood as indicating different origins for the comets, or at least different dynamical evolutions before they become observable. This is confirmed by the most recent studies (even though the cometary populations actually overlap in the space of orbital elements, implying that the period is not a good criterion for defining them, see e.g. \citealp{NESVORNY-etal_2017}).
   
   Revisiting an early concept exposed by \cite{OPIK_1932}, \cite{OORT_1950} showed that long-period comets come from a distant, roughly spherical reservoir, extending up to the very limit of the gravitational influence of the sun in its stellar environment. This reservoir, now called the Oort cloud, has been created by small bodies that were scattered away by the planets during the early stages of the formation of the solar system. At such large distances from the sun, Oort cloud bodies are subject to the gravitational torques caused by the overall galactic field (as it was understood by \citealp{HEISLER-TREMAINE_1986}), and from sporadic close passages of massive objects, like stars and molecular clouds. Upon the action of such external forces, bodies naturally spread into the isotropic distribution of long-period comets \citep{DUNCAN-etal_1987}. At the time of Oort's publication, no reservoir of small bodies was known apart from the main asteroid belt. Hence, Oort considered that main-belt asteroids could be the initial source of such scattered bodies. Shortly after Oort's work about long-period comets, \cite{KUIPER_1951} conjectured the existence of a belt of icy bodies lying beyond the orbit of Neptune and up to $50$~astronomical units (au). A similar conclusion had actually been drawn by \cite{EDGEWORTH_1949} a few years earlier, but neither Oort nor Kuiper were aware of his work. This icy belt, now called the Kuiper (or Edgeworth-Kuiper) belt, appeared in both studies as a natural consequence of the planetary formation process: since the outer edge of the protoplanetary disc should have been much slower and less dense than the rest of the disc, only small bodies could possibly have been formed there. Kuiper concluded that a fraction of icy bodies contained in this belt should have been scattered away by Pluto (which was thought at that time to be quite massive) and ejected into the Oort cloud. Indeed, the Kuiper belt, if it ever existed, was a more likely source of comets than the main belt, since the latter is mostly composed of rocky bodies, whereas comets were recently recognised by \cite{WHIPPLE_1950} to be icy.
   
   The source and the very origin of long-period comets seemed to be solved. But what about short-period comets? According to the scenario of \cite{KUIPER_1951}, the early scattering event that led to the formation of the Oort cloud should also have injected comets into the inner solar system. However, since the observed short-period comets are still active (i.e. they have not lost all of their icy content yet due to repeated passages near the sun), and since they have strongly unstable orbits, a steady supply of fresh comets was needed. From a long time \citep{TISSERAND_1889,CALLANDREAU_1892,NEWTON_1893}, it was known that nearly parabolic comets can be ``captured'' onto short-period orbits by repeated interactions with Jupiter. For decades after the hypothesis of \cite{OORT_1950}, astronomers tackled the problem of reproducing the observed low-inclination distribution of short-period comets from an isotropic reservoir of long-period comets. It was found that hundreds of perihelion passages were generally required to produce an orbit similar to those observed, with a high probability of ejection, pointing towards a very low-efficiency mechanism. Yet, low-inclinations comets are perturbed most due to their low encounter velocity with the planets; this produces a higher capture probability which was in apparent agreement with the observed distribution of short-period comets \citep{EVERHART_1972}. However, this paradigm definitely changed when \cite{FERNANDEZ_1980}, breathing new life into the original idea of \cite{EDGEWORTH_1949}, proved that, still today, a steady flux of short-period comets could be injected in the planetary region directly from the Kuiper belt. Due to its much higher efficiency rate, this mechanism would then produce most of the short-period comets, whose low inclinations would result from their initial disc-like distribution. This scenario was supported by the numerical simulations of \cite{DUNCAN-etal_1988}, strongly favouring the existence of this still-unobserved Kuiper belt. In 1992, the first object beyond Pluto was discovered \citep{JEWITT-LUU_1993}. Its orbit was roughly circular and barely inclined, as predicted for the Kuiper belt members. The following years, numerous other Kuiper belt objects were discovered. Further numerical experiments by \cite{LEVISON-DUNCAN_1997} revealed that, on their way towards becoming Jupiter-family or long-period comets, many unstable Kuiper belt objects remained wandering about chaotically in a long-lived transient state beyond Neptune. From this result, \cite{DUNCAN-LEVISON_1997} concluded that such long-lived scattering small bodies should have been produced in large quantities at the early stages of the formation of the solar system (i.e. when the Oort cloud has been formed, see above) and remain today in the form of a ``scattered disc''. This prediction almost coincided with its observational confirmation, since the first body recognised as a member of the scattered disc was discovered the same year \citep{LUU-etal_1997}. The scattered disc differs from the Kuiper belt as it was initially imagined by a broader distribution of eccentricity and inclination and by the unstable nature of its members. It is recognised today as producing the large majority of Jupiter-family comets (even though a fraction of them does come from the classical Kuiper belt, and a few from the Oort cloud). The scattered disc also produces a few Halley-type comets, but recent studies show that Halley-type comets mainly come from the Oort cloud, including its flattened inner component that is responsible for their slightly anisotropic distribution \citep{NESVORNY-etal_2017}.
   
   This distinction between the inner and outer components of the Oort cloud is a natural consequence of the efficiency of external perturbations, which decreases for decreasing heliocentric distances. Various dynamical arguments concur to place the limit at a semi-major axis of about $20\,000$~au. Most long-period comets are observed to come from the outer Oort cloud, where perturbations are the strongest, but this does not mean that the inner Oort cloud is empty. This was pointed out by \cite{HILLS_1981}, who predicted the existence of a very massive inner Oort cloud (or ``Hills cloud'') that could replenish the outer Oort cloud with a fraction of its lost comets. From a very different argument, \cite{LEVISON-etal_2001} found that a very massive inner Oort cloud was indeed required in order to match the low-inclination concentration of Halley-type comets: due to the weaker external perturbations, the inner Oort cloud still keeps today a clear memory of its initial disc-like distribution; it is therefore concentrated near the ecliptic, contrary to the isotropic outer Oort cloud (see e.g. \citealp{FOUCHARD-etal_2017,FOUCHARD-etal_2018}). However, as the number of observed Halley-type comets grew, their median inclination grew as well, weakening the need for a massive inner Oort cloud. The low-inclination distribution of Halley-type comets is now understood to be a statistical bias of previously incomplete datasets \citep{WANG-BRASSER_2014}. Modern simulations rather predict a similar number of objects in the inner and outer components of the Oort cloud, which equally contribute to the flux of Halley-type comets (see e.g. \citealp{NESVORNY-etal_2017,VOKROUHLICKY-etal_2019}). As we will see throughout this review, however, there is a continuous transfer of objects between the different reservoirs of small icy bodies, and the question of their origin somewhat loses its meaning (see also \citealp{LEVISON-etal_2006}).
   
   In parallel to the search for the origin of short-period comets, a new class of small bodies was recognised by  \cite{KOWAL-etal_1979} after the discovery of object (2060) Chiron. Indeed, such bodies have orbits much more eccentric and inclined than asteroids, but are larger than comets. They were called ``centaurs'' in reference to their property of being neither completely asteroids nor comets. This duality was confirmed with the discovery of their unexpected cometary activity \citep{HARTMANN-etal_1990}. These objects mostly have unstable orbits lying between Jupiter and Neptune. They are now known to be in a transitional orbital state between short-period comets and their reservoirs (scattered disc, Oort cloud, see above).
   
   As new trans-Neptunian objects were discovered, it appeared that their large variety of orbits cannot simply be divided into the Kuiper belt, the scattered disc and the Oort cloud. Classification problems reached their climax in 2004, with the discovery of Sedna \citep{BROWN-etal_2004}. Sedna was the first object discovered that appeared to be out of reach of any known orbital perturbation, and yet, it has a very eccentric orbit incompatible with an in-situ formation. Although the orbits of Sedna-like bodies continue to puzzle astronomers, current models of the formation of the solar system can explain their existence. After having detailed the main dynamical mechanisms at play beyond Neptune (Sects.~\ref{sec:adiff} to \ref{sec:stars}), the current classification of trans-Neptunian objects is given in Sect.~\ref{sec:sculpt}, along with our understanding of their origins.
   
   In a large variety of works, although not all, the terms ``Kuiper belt'' now generically encompass all small bodies with orbits beyond Neptune that receive negligible perturbations from the galactic tides. This distinguishes them from Oort-cloud comets whose dynamics, mostly governed by galactic tides and passing stars, is qualitatively very different. As we will see, the limit between the two populations is actually quite fuzzy and extends in a semi-major axis range from about $500$ to $1600$~au. The terms ``trans-Neptunian objects'' are sometimes used as a synonym of ``Kuiper belt'' in its broader sense, thus implicitly excluding the Oort cloud. In this review article, we rather consider all small bodies with semi-major axis larger than Neptune's.
   
\section{Planetary scattering}\label{sec:adiff}
   As illustrated in Fig.~\ref{fig:zones}, the planetary scattering is triggered below some threshold of the perihelion distance. For nearly planet-crossing orbits, this scattering is due to close encounters within the Hill sphere of the giant planets, which radius is about $1$~au for Neptune. In this case, all orbital elements change according to {\"O}pik's theory \citep{CARUSI-etal_1990,VALSECCHI-etal_1997,VALSECCHI-etal_2000,VALSECCHI-etal_2003,VALSECCHI-etal_2018}. The scattering region, however, extends well beyond the limit of such close encounters with Neptune. Confirming the early results by \cite{TORBETT-SMOLUCHOWSKI_1990}, \cite{GLADMAN-etal_2002} showed that the scattering effect of Neptune is significant over long timescales for perihelion distances below about $45$~au. They also found that the precise limit actually increases with the semi-major axis value. The slope of this limit was further investigated by \cite{GALLARDO-etal_2012} up to high orbital inclinations; as detailed below, this slope can be qualitatively understood by simple considerations.
   
   In fact, some observed objects with perihelion beyond $45$~au are known to experience scattering \citep{BANNISTER-etal_2017}. Strictly speaking, the scattering of such distant trans-Neptunian objects is not due to close encounters with Neptune, even though Neptune is indeed the main responsible, and the main perturbations do happen at perihelion owing to the very large eccentricity of these bodies. Instead, this scattering comes from an overlap of mean-motion resonances with the giant planets, mostly with Neptune. For extreme eccentricities, the resonance widths are very large almost independently of the resonance order (see Sect.~\ref{sec:res}), leading to a massive overlap. According to Chirikov's criterion \citep{CHIRIKOV_1960}, the momentum conjugate to the mean longitude of the small body (i.e. its semi-major axis) suffers from stochastic jumps that are localised inside the region of overlap: this is the essence of planetary scattering. In the limiting case of close encounters with the planets, the overdensity of overlapping resonances naturally generates the kick of {\"O}pik's theory, in a similar way as the Dirac $\delta$ function can be constructed from an infinite sum of cosine harmonics.
   
   Additionally to mean-motion resonances, the scattering region beyond Neptune contains a few secular resonances \citep{KNEZEVIC-etal_1991,DUNCAN-etal_1995,MORBIDELLI-etal_1995}. However, considering the very slow orbital precession of trans-Neptunian objects (see Sect.~\ref{sec:sec}), these secular resonances are restricted to small semi-major axes, not larger than $50$~au. We will therefore ignore secular resonances in the discussions below.
   
   Since most of the orbital perturbations occur at perihelion, the perihelion distance $q$ of the small body remains almost unchanged during the process of planetary scattering \citep{DUNCAN-etal_1987}. Moreover, since Neptune is the main perturber and that its orbit is almost circular, the Tisserand parameter with respect to the three-body problem Sun-Neptune-body is also approximately constant \citep{TISSERAND_1889b}:
   \begin{equation}
      T = \frac{a_\mathrm{N}}{a} + 2\sqrt{\frac{a}{a_\mathrm{N}}(1-e^2)}\cos I \,.
   \end{equation}
   In this expression $a_\mathrm{N}$ is the semi-major axis of Neptune, $a$ is the semi-major axis of the small body, $e$ its eccentricity, and $I$ its inclination. In the limit of far-away scattering, $a$ tends to infinity while $q$ remains close to Neptune, simplifying the Tisserand parameter to $T=2\sqrt{2q/a_\mathrm{N}}\cos I$. This expression shows that $I$ cannot vary much since $q$ and $T$ are almost constant during the scattering process. Actually, for an orbit initially circular and lying in the ecliptic scattered away by Neptune, the largest possible inclination reachable is about $30^\text{o}$. This number is obtained assuming the lowest possible value of $T$ (namely, $3$) and the most efficient chaotic diffusion for both $q$ (raised all the way up to $45$~au) and $a$ (sent to infinity). This limit is quite extreme, and pure scatterers have a very low probability of reaching it in a timespan restricted to the age of the solar system \citep{GOMES_2003,LYKAWKA-MUKAI_2007}. However, other mechanisms can contribute to raise the inclination and perihelion distance of trans-Neptunian objects, like isolated mean-motion resonances (Sect.~\ref{sec:res}) or galactic tides (Sect.~\ref{sec:gt}).
   
   There are two ways of examining the scattering process. One way is to first consider each relevant resonance individually, and to locate the regions where they overlap. Chaotic diffusion coefficients can then be estimated \citep{MURRAY-etal_1985,MURRAY-HOLMAN_1997}, as well as the limits of the chaotic region. In Sect.~\ref{sec:res}, we will present a semi-analytical method that can be used to measure the width of any mean-motion resonance at first order of the planetary perturbation, for any value of the eccentricity and inclination, as done for instance by \cite{MORBIDELLI-etal_1995}. However, as usual when using Chirikof's criterion, the chaos appears a little before the predicted limit because of the overlap of higher-order resonances (i.e. resonances that appear at higher order in the Hamiltonian developed in Lie series, such as three-body resonances). The structure of the limit is actually fractal-like, with always higher and higher-order resonances to be taken into account in order to better resolve the limit (see e.g. the maps by \citealp{ROBUTEL-LASKAR_2001}). This fractal structure justifies the use of numerical methods for getting accurate estimates of the limit of the chaotic region (i.e. the boundary between the scattered and detached populations described by \citealp{LYKAWKA-MUKAI_2007}).
   
   The second way of examining the scattering process is to consider that at each revolution, the small body receives a kick from the planets. This method is efficient for very eccentric bodies, because they mostly follow unperturbed Keplerian orbits around the barycentre of the solar system, and only feel the planetary perturbations at perihelion during a very short amount of time. Analytical and semi-analytical estimates of the kicks can therefore be computed and used to map the orbital evolution from one perihelion passage to the next one (see e.g. \citealp{MALYSHKIN-TREMAINE_1999}, \citealp{PAN-SARI_2004}, \citealp{FOUCHARD-etal_2013}, and the review by \citealp{SHEVCHENKO_2011}). On these maps, the stable resonant regions are easily localised, as well as the fractal-like border of the chaotic region. Lyapunov exponents can also be estimated as a measure of the chaotic diffusion \citep{SHEVCHENKO_2007}. As shown by \cite{PAN-SARI_2004}, the energy kicks received at each perihelion passage are almost independent on the semi-major axis $a$. They strongly depend on $q$, however, and become vanishingly small beyond some value of $q$. This is why there is no planetary scattering in the top portion of Fig.~\ref{fig:zones}. Moreover, energy kicks are almost symmetrically distributed around zero (the symmetry is exact in the planar case and at first order to the planetary perturbations, see \citealp{PAN-SARI_2004}). At each perihelion passage, the small body has therefore an equal probability of receiving a positive or a negative energy kick. From these basic properties, fundamental characteristics of the planetary scattering can be understood. Introducing the variable $z=-1/a$, which is proportional to the orbital energy of the small body, we note that an energy kick $\Delta z$ produces a net semi-major axis variation of
   \begin{equation}\label{eq:kick}
      \Delta a = \frac{a^2\Delta z}{1-a\Delta z} \,.
   \end{equation}
   Typical values for $\Delta z$ can be found in \cite{FOUCHARD-etal_2013} in the limit of large semi-major axes. For a zero-inclination orbit, $|\Delta z|$ is smaller than $10^{-6}$~au$^{-1}$ for $q\gtrsim 55$~au, but it rapidly increases for decreasing perihelion distance. For $q\approx 30$~au, $|\Delta z|$ can exceed $0.1$~au$^{-1}$. However, the net value of $\Delta z$ strongly depends on the geometry of the encounter with Neptune; more details about the kick function can be found in \cite{MALYSHKIN-TREMAINE_1999} or \cite{PAN-SARI_2004}.
   
\begin{figure}
      \centering
      \includegraphics[width=\textwidth]{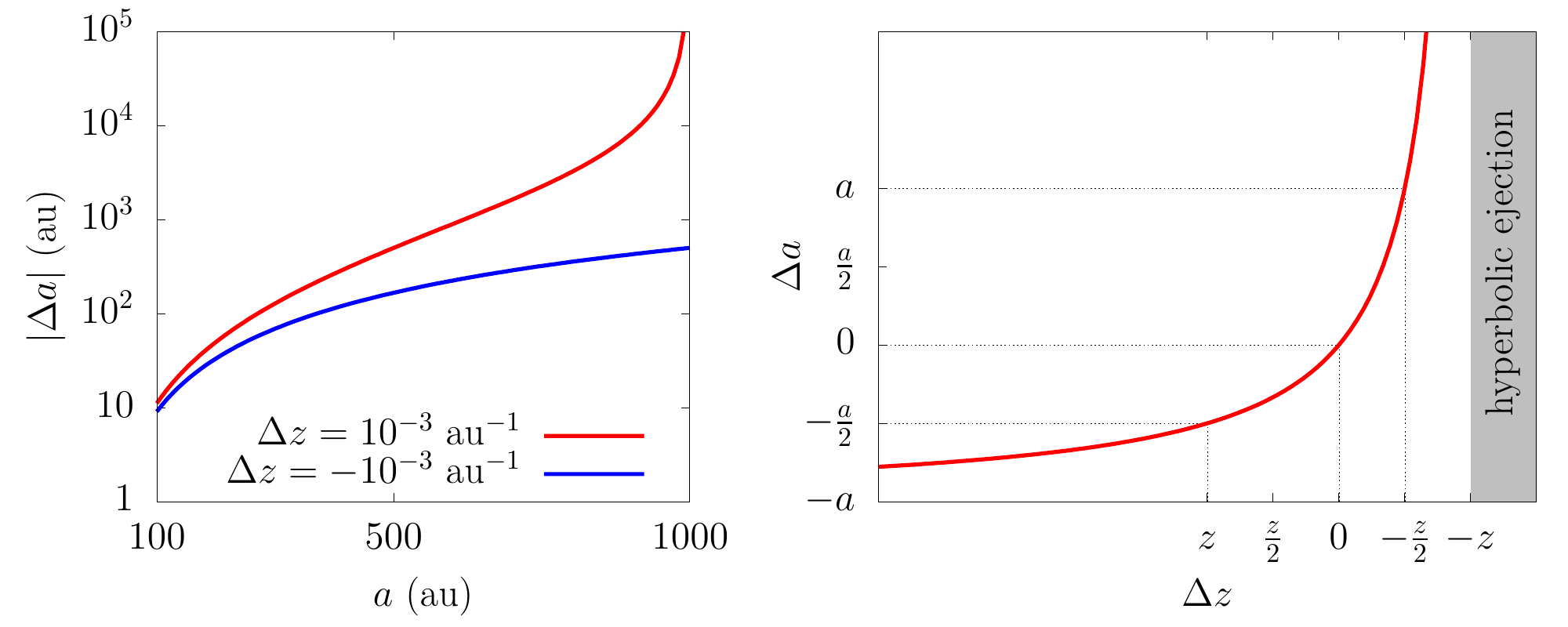}
      \caption{Variation of semi-major axis as a function of the energy kick. \emph{Left:} $\Delta a$ as a function of $a$ for a fixed energy kick taken as example. The red curve tends to infinity at $a = 1000$~au (parabolic ejection). The blue curve represents a negative $\Delta a$. \emph{Right:} $\Delta a$ as a function of $\Delta z$ for a fixed semi-major axis $a$. The curve tends to infinity at $\Delta z = -z$ (parabolic ejection) and to $-a$ for $\Delta z\rightarrow -\infty$.}
      \label{fig:kick}
   \end{figure}

\begin{figure}
      \centering
      \includegraphics[width=0.8\textwidth]{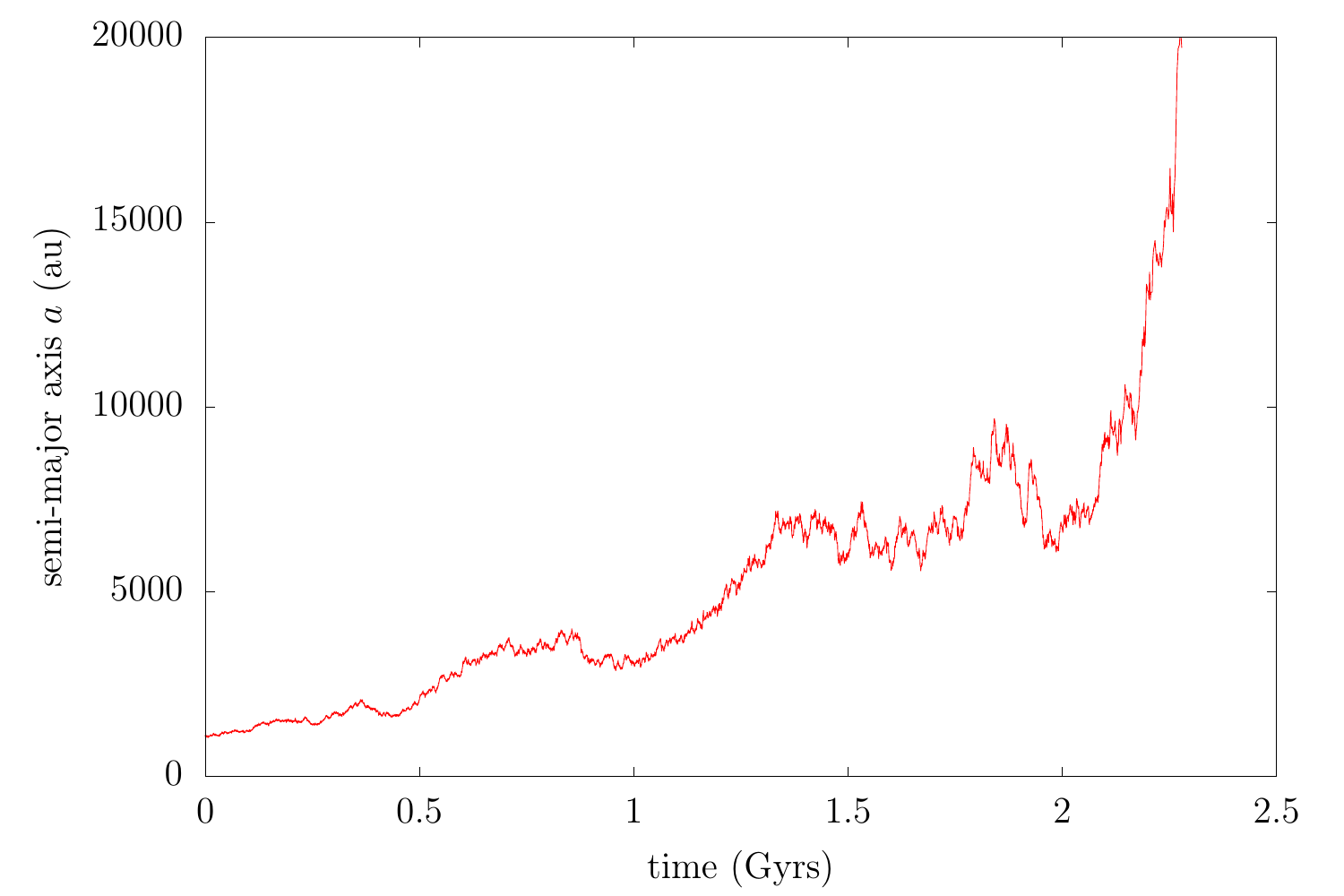}
      \caption{Ejection of a scattered-disc object obtained by numerical integration. The system contains the sun, the four giant planets starting on their current orbits, and the small body. The initial conditions of the small body are $a=1000$~au, $q=47$~au and $I=17^\text{o}$. A high initial perihelion distance is chosen in order to slow down the scattering, and galactic tides are not included. During the whole simulation, $q$ and $I$ vary by $0.6$~au and $2^\text{o}$, respectively.}
      \label{fig:ejection}
   \end{figure}

   Equation~\ref{eq:kick} shows that for a given energy kick, the resulting variation of semi-major axis is much larger for larger $a$, as illustrated in the left panel of Fig.~\ref{fig:kick}. Moreover, negative energy kicks are much less efficient in reducing $a$ than positive energy kicks are in increasing $a$, as illustrated in the right panel of Fig.~\ref{fig:kick}. This asymmetry indicates that the limiting value of $q$ above which the diffusion can be neglected is not constant with $a$, but slightly increases, as illustrated in Fig.~\ref{fig:zones}. On the limit, negative energy kicks are small enough to be neglected, but not positive ones. Determining the slope of this limit is a complex task. Based on numerical experiments, empirical expressions of the limit were given by several authors. In the range $a\in[30,90]$~au and for $I=0$, \cite{ROBUTEL-LASKAR_2001} gave the expression $q=0.196\,a + 30$~au. In the range $a\in[30,270]$~au and for $I=17^\text{o}$, \cite{GLADMAN-etal_2002} then gave the expression $q=0.09\,a+27.3$~au. In the range $a\in[30,700]$~au and for various values of $I<70^\text{o}$, \cite{GALLARDO-etal_2012} finally gave the expression $q=0.037\,a+33.3$~au. Even though the orbital inclination does play a role in fixing the limit, the large differences between these estimates reflect the level of arbitrariness inherent to the classification between the presence or absence of chaotic diffusion. As we already mentioned, the real limit is actually fractal-like and its structure cannot be understood without going much more into dynamical details. Based on the qualitative effects of energy kicks, however, one can understand the runaway diffusion observed in numerical simulations, that often leads to ejection of small bodies. Indeed, since $q$ is almost unaffected by the scattering process, the energy kicks remain roughly equally large if $a$ grows, but lead to larger variations of $a$, that are even larger if they are positive again (see Eq.~\ref{eq:kick} and Fig.~\ref{fig:kick}). Consequently, the increase of $a$ can get faster and faster until the energy becomes positive. An example of such an evolution is given in Fig.~\ref{fig:ejection}: even though the energy kicks approximately keep the same distribution, the jumps of $a$ get clearly bigger as $a$ increases, until the object is brutally ejected.
   
\begin{figure}
      \centering
      \includegraphics[width=\textwidth]{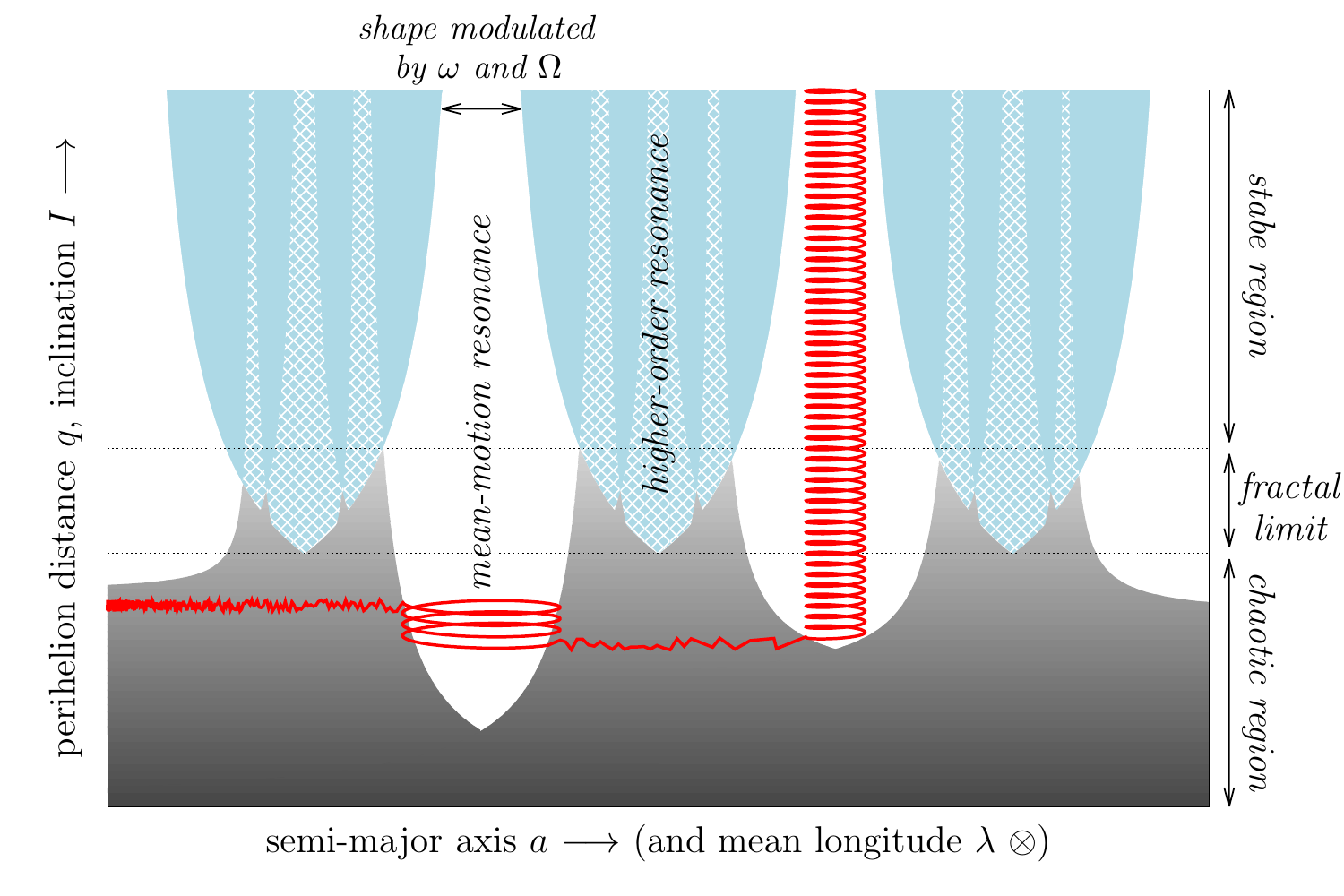}
      \caption{Schematic representation of the scattering region and its connection to the inert zone. This is a zoom-in view of the bottom left portion of Fig.~\ref{fig:zones} ($a\sim[80,500]$~au and $q\sim[30,80]$~au). The mean longitude $\lambda$ (in some rotating frame) is along the axis perpendicular to the plane of the figure. The chaotic scattering region is represented in grey, and the stable non-overlapping portions of the resonances are coloured white, with more transparency for resonances appearing at second and higher order of the perturbations. Inside the chaotic volume, energy kicks are strong in the bottom (dark shades), and weak in the top (light shades). The red curve represents an example of trajectory of a scattered small body. Starting on the left-hand side, the semi-major axis first diffuses towards larger values. Then, the small body encounters a resonant tube, but the capture is only temporary because the resonant dynamics makes $q$ decrease, eventually releasing the small body in the chaotic volume again. The diffusion is reactivated, and it is faster than before because $a$ is larger (see text) and $q$ is smaller. Finally, the small body is captured in a second resonance which lifts it this time outside of the chaotic volume.}
      \label{fig:tubes}
   \end{figure}

   The precise fractal-like limit of the region of planetary scattering in the whole space of orbital elements is still not fully characterised, even though major advances have been made semi-analytically by \cite{MORBIDELLI-etal_1995} and D.~Nesvorn{\'y}\footnote{Unpublished results described in the web page:\\ \texttt{https://www.boulder.swri.edu/\textasciitilde{}davidn/kbmmr/kbmmr.html}.} in terms of resonance overlap, and numerically by \cite{ROBUTEL-LASKAR_2001} in terms of chaos mapping. In order to investigate the dynamical structure in the vicinity of specific resonances, many other authors turned to purely numerical methods \citep{LEVISON-STERN_1995,MORBIDELLI_1997,NESVORNY-ROIG_2000,NESVORNY-ROIG_2001,KOTOULAS-VOYATZIS_2004}. In the planar case, the reduced number of degrees of freedom allows one to study the structure of the scattering region using Poincar{\'e} surfaces of section. This was realised by \cite{MALHOTRA_1996}, \cite{MALHOTRA-etal_2018}, \cite{LAN-MALHOTRA_2019}, and by \cite{PAN-SARI_2004} using the kick formalism. Based on these previous works, and on results about the resonant dynamics (see Sect.~\ref{sec:res}), we can get a clear qualitative idea of the orbital dynamics inside the scattering region. As illustrated in Fig.~\ref{fig:tubes}, one can think of the scattering region as a chaotic volume in the space of orbital elements, where the horizontal plane is spanned by the semi-major axis $a$ and the mean longitude $\lambda$, and the vertical dimension corresponds to the perihelion distance $q$ and the inclination $I$ (we forget for now about the argument of perihelion $\omega$ and the longitude of node $\Omega$). When inside the chaotic volume, a small body evolves almost\footnote{For a slow scattering, the secular perturbations of the planets actually produce small oscillations of $q$ and $I$, especially near $I\approx 63^\text{o}$ and $117^\text{o}$, as described in Sect.~\ref{sec:sec}. Even without isolated mean-motion resonances, a small body can hence slowly diffuse vertically inside the chaotic volume.} in a horizontal sheet, with the asymmetric random walk of $a$ due to energy kicks. However, the chaotic volume is pierced with vertical porous tubes of stability corresponding to the non-overlapping portions of the mean-motion resonances with the planets (see the maps by \citealp{ROBUTEL-LASKAR_2001}, which are upside-down with respect to Fig.~\ref{fig:tubes} because their vertical axis shows $e$ instead of $q$). At their bottom extremity, these vertical tubes are immersed inside the chaotic volume. The shape and width of the tubes are modulated by the value of $\omega$ and $\Omega$, which are made to circulate because of the secular action of the planets (see Sect.~\ref{sec:sec}). As measured by \cite{LAN-MALHOTRA_2019}, the resonant tubes (i.e. the stable portions of the resonances) cover quite a substantial fraction of the chaotic volume. Wandering about on its horizontal sheet, the small body has therefore a reasonable probability of encountering a resonant tube, and whenever it comes with suitable phases for $\omega$ and $\Omega$, the resonant dynamics absorbs it inside the tube. In this case, the diffusion of $a$ stops and a combination of mean longitudes starts to oscillate: this is a capture in mean-motion resonance.
   
   Since it does not involve any dissipative process, such a resonant capture is always reversible. As detailed in Sect.~\ref{sec:res}, there are many pathways inside a resonant tube. Most of them simply bring back the small body towards the chaotic volume on a slightly different horizontal sheet, producing a vertical migration (see e.g. \citealp{LYKAWKA-MUKAI_2007b}). This mechanism was described quite early by \cite{DUNCAN-LEVISON_1997}, even though its precise nature was poorly understood at that time. For instance, a release out of resonance happens if the variations of $\omega$ and $\Omega$ narrow the tube, or if $q$ decreases, leaving the small body outside again in the chaotic volume. A few pathways, however, lift the small body very high in perihelion distance, completely above the chaotic volume, inside the region labelled ``isolated planetary resonances'' in Fig.~\ref{fig:zones}. Small bodies can remain there for time periods longer than the age of the solar system. Their fate is described in Sect.~\ref{sec:res}.
   
   Beyond the limit fixed at $a\approx 500$~au in Fig.~\ref{fig:zones}, the resonant tubes are not wide enough to guarantee stable captures, and the small bodies cannot substantially migrate in the vertical direction, let alone escape the chaotic volume, through this mechanism. As shown in Fig.~\ref{fig:zones}, however, if they acquire a large-enough semi-major axis without being ejected on a hyperbolic orbit, the galactic tides come into play and are able to extract them, at least temporarily, from the scattering region \citep{DUNCAN-etal_1987,LEVISON-etal_2006}. When small bodies cycle back towards the scattering region, the galactic tides may have given them large orbital inclinations that do not drop back to their original values, contrary to the perihelion distance \citep{HIGUCHI-etal_2007}. As detailed further in Sect.~\ref{sec:gt}, this mechanism can therefore produce scattered-disc objects ($q>30$~au) or centaurs ($q<30$~au) with very high inclinations \citep{KAIB-etal_2019}, as well as Halley-type comets \citep{LEVISON-etal_2006}.
   
   Little is known about planetary scattering at high orbital inclinations, given that very few trans-Neptunian objects have $I>50^\text{o}$, as expected from formation scenarios. Even though the picture described previously is known to remain qualitatively correct, the limits of the chaotic region, the diffusion timescales, and the structure of the resonances are expected to differ substantially. The map of \cite{ROBUTEL-LASKAR_2001} computed for $I=30^\text{o}$ already shows a departure from the planar case: the chaotic region is restricted to smaller perihelion distances and the stable resonant zones appear to be more numerous. This is somehow confirmed by the numerical explorations of \cite{GALLARDO_2019b}, even though they are restricted to $a<38$~au, revealing a large stability region located at $I\approx 150^\text{o}$. Further investigations would be required to determine the nature of this region, its extent for larger semi-major axes, and the dynamics in its vicinity.
   
\section{Secular dynamics driven by the giant planets}\label{sec:sec}
   In this section, we focus on the regions of Fig.~\ref{fig:zones} where the planetary scattering is inefficient and the orbital dynamics is dominated by the so-called secular planetary perturbations. This mostly concerns the blue zones tagged as ``inert'': between the isolated mean-motion resonances and beyond their reach. The reason for this inactivity will become clearer by going a little further into details about the dynamics. The heliocentric Keplerian elements of the small body are written $(a,e,I,\omega,\Omega,M)$, with the associated canonical Delaunay elements:
   \begin{equation}\label{eq:Del}
      \left\{
      \begin{aligned}
         \ell &= M \\
         g &= \omega \\
         h &= \Omega
      \end{aligned}
      \right.
      \hspace{0.5cm},\hspace{0.5cm}
      \left\{
      \begin{aligned}
         L &= \sqrt{\mu a} \\
         G &= \sqrt{\mu a(1-e^2)} \\
         H &= \sqrt{\mu a(1-e^2)}\cos I
      \end{aligned}
      \right.\,,
   \end{equation}
   where $\mu$ is the gravitational parameter of the sun. The Hamiltonian governing the orbital motion of the small body around the sun perturbed by $N$ planets can be written
   \begin{equation}\label{eq:Hfirst}
      \mathcal{H} = \mathcal{H}_0(L) + \varepsilon_\mathrm{P}\mathcal{H}_\mathrm{P}(L,G,H,\ell,g,h,t) \,,
   \end{equation}
   where $\varepsilon_\mathrm{P}\ll 1$ and
   \begin{equation}
      \left\{
      \begin{aligned}\label{eq:H0H1}
         \mathcal{H}_0 &= -\frac{\mu^2}{2L^2} \,,\\
         \varepsilon_\mathrm{P}\mathcal{H}_\mathrm{P} &= - \sum_{i=1}^{N}\mu_i\left(\frac{1}{\|\mathbf{r}-\mathbf{r}_i\|} - \mathbf{r}\cdot\frac{\mathbf{r}_i}{\|\mathbf{r}_i\|^3}\right) \,.
      \end{aligned}
      \right.
   \end{equation}
   Here, P stands for ``planets''. The vectors $\mathbf{r}$ and $\mathbf{r}_i$ are the heliocentric position of the small body and of planet $i$, and $\mu_i$ is the gravitational parameter of planet $i$. The Hamiltonian explicitly depends on the time $t$ through the positions $\{\mathbf{r}_i\}$ of the $N$ planets. Seen from distant trans-Neptunian objects, the planetary orbits are very well approximated by circular and coplanar trajectories (see e.g. \citealp{THOMAS-MORBIDELLI_1996}, \citealp{SAILLENFEST-etal_2016}). Consequently, we set $\mathbf{r}_i = a_i(\cos\lambda_i,\sin\lambda_i,0)^\mathrm{T}$, where $a_i$ is constant, and we consider that all the higher-order terms in planetary eccentricities and inclinations are of order $\mathcal{O}(\varepsilon_\mathrm{P}^2)$ and can be neglected. In this case, the time $t$ can be replaced by the $N$ mean longitudes $\{\lambda_i\}$, with conjugate momenta $\{\Lambda_i\}$. The Hamiltonian function is then made autonomous by redefining $\mathcal{H}_0$:
   \begin{equation}
      \mathcal{H}_0 = -\frac{\mu^2}{2L^2} + \sum_{i=1}^Nn_i\Lambda_i \,,
   \end{equation}
   where $n_i$ is the mean motion of planet $i$, related to $a_i$ through $n_i^2a_i^3 = \mu + \mu_i$. In the regions where the diffusion of semi-major axis is inefficient (see Sect.~\ref{sec:adiff}), the orbit of a trans-Neptunian object is subject to fast periodic changes (frequency $\propto 1$) and a long-term modulation (frequency $\propto \varepsilon_\mathrm{P}$). In this case, the dynamics is better described in coordinates that are averaged over the short-period terms, called ``secular coordinates''. Assuming that the small body is out of any mean-motion resonance with the planets, the Hamiltonian function in the new coordinates can be developed in Lie series, as
   \begin{equation}\label{eq:Fsec}
      \mathcal{M} = \mathcal{M}_0(L) + \varepsilon_\mathrm{P}\mathcal{M}_\mathrm{P}(L,G,H,g) + \mathcal{O}(\varepsilon_\mathrm{P}^2) \,,
   \end{equation}
   where
   \begin{equation}\label{eq:F0F1}
      \left\{
      \begin{aligned}
         \mathcal{M}_0 &= -\frac{\mu^2}{2L^2} + \sum_{i=1}^Nn_i\Lambda_i \,, \\
         \varepsilon_\mathrm{P}\mathcal{M}_\mathrm{P} &= -\sum_{i=1}^{N}\frac{1}{4\pi^2}\int_{0}^{2\pi}\!\!\!\int_{0}^{2\pi}\!\!\!\frac{\mu_i}{|\mathbf{r}-\mathbf{r}_i|}\,\mathrm{d}\lambda_i\,\mathrm{d}\ell \,.
      \end{aligned}
      \right.
   \end{equation}
   Even though we use the same symbols as before, the coordinates are now the secular ones. Since the indirect part of the planetary perturbations vanishes over the average, it was omitted in Eq.~\eqref{eq:F0F1}. At first order in $\varepsilon_\mathrm{P}$, since the secular Hamiltonian function $\mathcal{M}$ is independent of $\ell$, the secular momentum $L$ (and hence the secular semi-major axis) is a constant of motion. The same holds for each $\Lambda_i$. Moreover, as noted by \cite{LIDOV_1962} and \cite{KOZAI_1962}, the rotational symmetry of the perturbation results in the constancy of $H$ as well, because the Hamiltonian function is also independent of $h$. We are left with a one-degree-of-freedom Hamiltonian system with two parameters: the semi-major axis $a$ and the Kozai constant
   \begin{equation}\label{eq:K}
      K = \sqrt{1-e^2}\cos I \,.
   \end{equation}
   The constancy of $K$ results in an exchange between eccentricity and inclination. Moreover, the inclination cannot cross the $90^\text{o}$ limit for $e<1$, otherwise $K$ would change sign.
   
\begin{figure}
      \centering
      \includegraphics[width=0.49\textwidth]{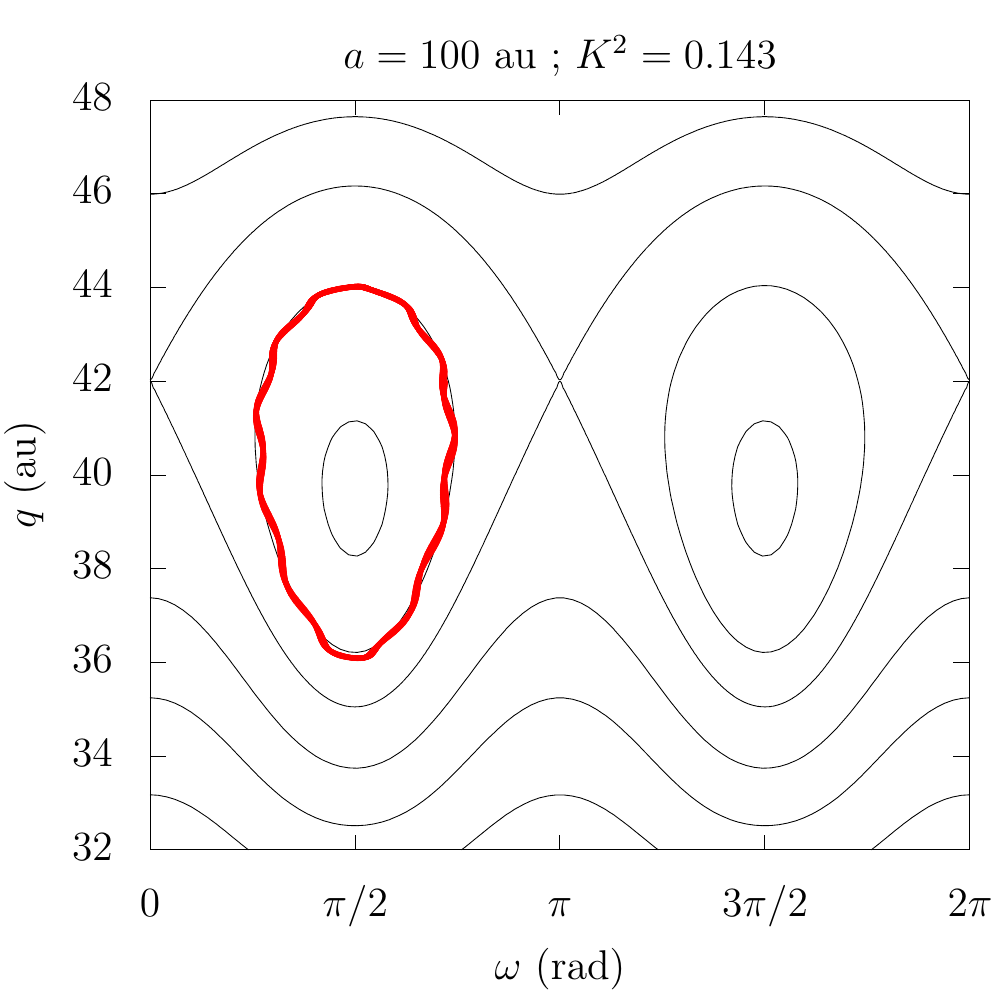}
      \includegraphics[width=0.49\textwidth]{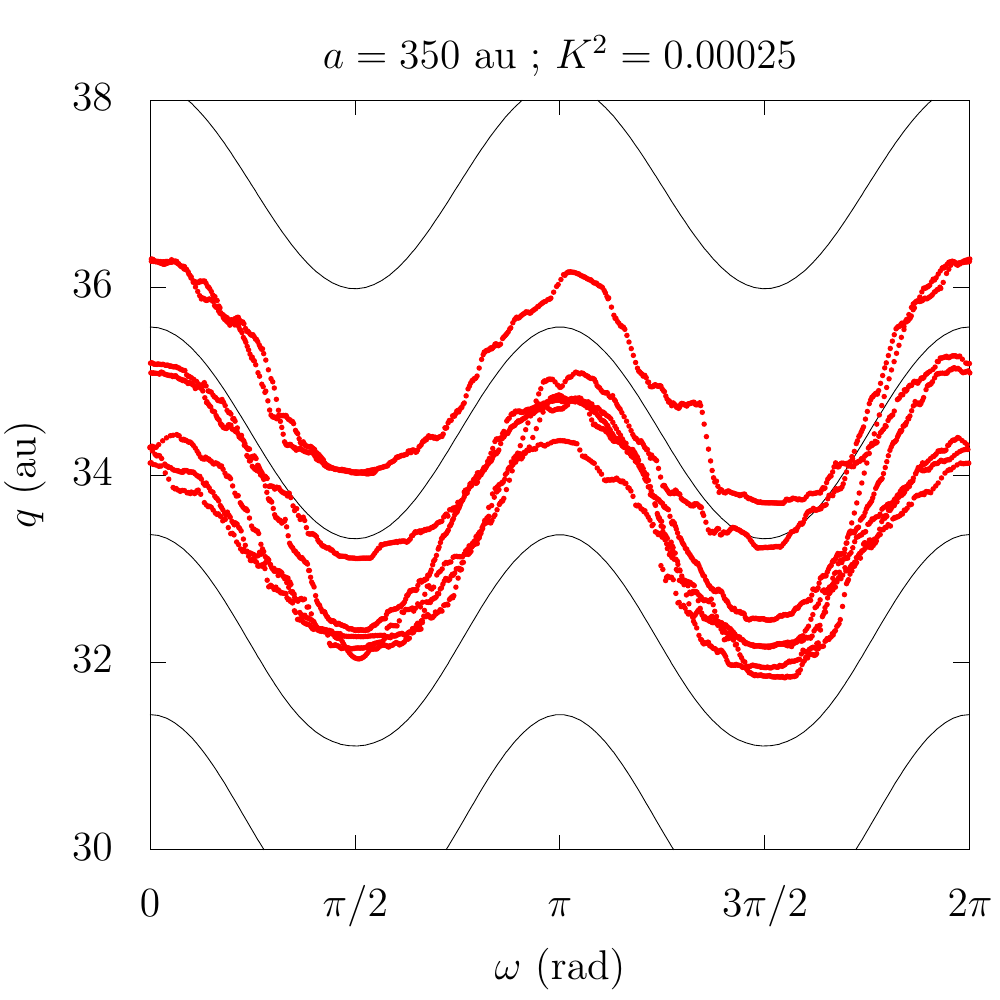}
      \caption{Unaveraged numerical integration compared to the secular dynamics. The red points represent two direct integrations for $4.5$~Gyrs (Hamiltonian $\mathcal{H}$ from Eq.~\ref{eq:Hfirst}), where the orbital elements of the planets vary according to the synthetic representation of \cite{LASKAR_1990}. The trajectory is plotted in barycentric coordinates against the level curves of the numerically-computed secular Hamiltonian $\mathcal{M}$ from Eq.~\eqref{eq:F0F1} with parameters indicated above the figures. (Strictly speaking, the Hamiltonian $\mathcal{M}$ is defined in heliocentric coordinates, but since we use secular elements, the wobbles of the sun are averaged out, so that barycentric and heliocentric coordinates become equivalent.) \emph{Left:} The semi-major axis is very stable at $a\approx 100$~au, and $\omega$ oscillates inside the libration island located at $I\approx 117^\text{o}$. \emph{Right:} The semi-major axis diffuses slowly between $250$ and $450$~au, with an orbital inclination close to $90^\text{o}$. In both cases, a high orbital inclination is chosen in order to limit the scattering (see Sect.~\ref{sec:adiff}).}
      \label{fig:Kozai}
   \end{figure}

   Neglecting $\mathcal{O}(\varepsilon_\mathrm{P}^2)$, the dynamics with Hamiltonian function $\mathcal{M}$ given by Eq.~\eqref{eq:Fsec} has been extensively studied in many situations (\citealp{THOMAS-MORBIDELLI_1996}, \citealp{GALLARDO-etal_2012}, \citealp{SAILLENFEST-etal_2016}). For objects with trajectories entirely exterior to the planetary orbits, it is convenient to develop $\varepsilon_\mathrm{P}\mathcal{M}_\mathrm{P}$ in Legendre polynomials, resulting in series of the semi-major axis ratios. This way, the truncated expression remains valid for any value of the eccentricity and inclination, and the development converges very quickly for distant objects. We obtain
   \begin{equation}
      \varepsilon_\mathrm{P}\mathcal{M}_\mathrm{P} = \varepsilon_{\mathrm{P}_0}\mathcal{M}_{\mathrm{P}_0} + \varepsilon_{\mathrm{P}_2}\mathcal{M}_{\mathrm{P}_2} + \varepsilon_{\mathrm{P}_4}\mathcal{M}_{\mathrm{P}_4} + \mathcal{O}(\varepsilon_{\mathrm{P}_6})\,,
   \end{equation}
   where these terms correspond to the monopole (index 0), quadrupole (index 2) and hexadecapole (index 4) of the expansion. The neglected terms are proportional to $(a_i/a)^6$. Writing
   \begin{equation}\label{eq:epsP}
      \varepsilon_{\mathrm{P}_0} = \frac{1}{a}\sum_{i=1}^N\mu_i
      \hspace{0.5cm},\hspace{0.5cm}
      \varepsilon_{\mathrm{P}_2} = \frac{1}{a^3}\sum_{i=1}^N\mu_ia_i^2
      \hspace{0.5cm},\hspace{0.5cm}
      \varepsilon_{\mathrm{P}_4} = \frac{9}{16}\frac{1}{a^5}\sum_{i=1}^N\mu_ia_i^4 \,,
   \end{equation}
   the explicit expression of each term is
   \begin{equation}\label{eq:Fp}
      \left\{
      \begin{aligned}
         \mathcal{M}_{\mathrm{P}_0} &= -1 \,,\\
         \mathcal{M}_{\mathrm{P}_2} &= \frac{1-3\cos^2I}{8(1-e^2)^{3/2}} \,,\\
         \mathcal{M}_{\mathrm{P}_4} &= \frac{1}{64(1-e^2)^{7/2}}\Big( (2+3\,e^2)(-3+30\cos^2I-35\cos^4I) \\&
         \hspace{2.8cm}+ 10e^2(1-7\cos^2I)\sin^2I\cos(2\omega)\Big) \,.
      \end{aligned}
      \right.
   \end{equation}
   The dynamics is trivially integrable at quadrupole order, that is, neglecting $\mathcal{O}(\varepsilon_{\mathrm{P}_4})$. Indeed, the momentum $G$ is conserved, meaning that both the inclination and the eccentricity are constant. The angles precess at constant velocity:
   \begin{equation}\label{eq:omegadot}
      \dot{\omega} = \varepsilon_{\mathrm{P}_2}\frac{3(5\cos^2I-1)}{8\sqrt{\mu a}(1-e^2)^2}
      \hspace{0.3cm};\hspace{0.3cm}
      \dot{\Omega} = \varepsilon_{\mathrm{P}_2}\frac{-3\cos I}{4\sqrt{\mu a}(1-e^2)^2} \,.
   \end{equation}
   This precession is similar to the one induced by a $J_2$ oblateness of the central body. It rapidly slows down for growing $q$ and $a$, preventing any secular resonance with the planets beyond about $a=50$~au \citep{KNEZEVIC-etal_1991}. As an example, for $a=100$~au, $q=50$~au, and $I=0$, the precession periods of $\omega$ and $\Omega$ are approximately $30$ and $60$~Myrs, respectively, which are about $15$ and $35$ times the values for Neptune \citep{LASKAR_1990}. From Eq.~\eqref{eq:omegadot}, we note that $\omega$ increases for $I$ between $0$ and about $63^\text{o}$, decreases between $63^\text{o}$ and $117^\text{o}$, and increases again beyond $117^\text{o}$. In contrast, $\Omega$ decreases for $I<90^\text{o}$, and increases beyond. By computing the average of Eq.~\eqref{eq:F0F1} numerically, \cite{GALLARDO-etal_2012} showed that the hexadecapolar and higher-order terms make libration islands appear for $\omega$ at orbital inclinations near the critical values of $I=63^\text{o}$ and $I=117^\text{o}$. According to \cite{SAILLENFEST-etal_2016}, these islands have a maximum width of $16.4$~au in perihelion distance, which, for large semi-major axes, only represents a small interval of eccentricity. From the constancy of $K$ (see Eq.~\ref{eq:K}), this converts into a tiny interval of orbital inclination near the two critical values. We now understand why the regions located in between the isolated resonances in Fig.~\ref{fig:zones} are inert: the perihelion distance and inclination are almost fixed, except for $I\approx 63^\text{o}$ or $117^\text{o}$, where they undergo quite moderate variations. The orbit is only affected by a slow precession of $\omega$ and $\Omega$. The situation is thus very different from the classical Lidov-Kozai mechanism raised by an external perturber \citep{LIDOV_1962,KOZAI_1962}, for which huge orbital variations can occur. But still, Fig.~\ref{fig:Kozai} shows that this mechanism has an effect in the long-term dynamics of trans-Neptunian objects, even if the semi-major axis diffuses slowly.
   
   If at least one of the planets included in the system has an eccentric and/or inclined orbit, the rotational symmetry is broken and $K$ is not conserved anymore. The system has therefore more than one degree of freedom, and one must turn to numerical integrations of the secular system, in the spirit of \cite{GRONCHI_2002} or \cite{TOUMA-etal_2009}. This method has been used by \cite{SAILLENFEST-etal_2017b} when including the effects of the planet proposed by \cite{BATYGIN-BROWN_2016} in the secular Hamiltonian from Eq.~\eqref{eq:F0F1}.
   
\section{Mean-motion resonances with the giant planets}\label{sec:res}
   If the small body presents a mean-motion resonance with one of the planets (i.e. there is a commensurability between their two orbital periods), the change of coordinates to secular variables used in Sect.~\ref{sec:sec} is not defined anymore and some terms in the neglected part of Eq.~\eqref{eq:Fsec} become overly large. This phenomenon happens in the scattering region of Fig.~\ref{fig:zones} (overlap of mean-motion resonances), and in the high-perihelion zone labelled ``isolated planetary resonances''. Because of the high eccentricities reached by trans-Neptunian objects, very distant mean-motion resonances with Neptune have a strong influence on their dynamics (the notion of ``resonance order'' actually loses its meaning, see e.g. \citealp{PAN-SARI_2004}). Yet, despite extreme eccentricities, numerical simulations show that isolated mean-motion resonances with the planets become inefficient beyond some semi-major axis threshold (\citealp{GOMES-etal_2005,GALLARDO-etal_2012,SAILLENFEST-etal_2017a,NESVORNY-etal_2017,VOKROUHLICKY-etal_2019}). This threshold is not well defined, but it is most likely related to the resonance strength (\citealp{GALLARDO_2006b}), that steadily decreases with the distance. In Fig.~\ref{fig:zones}, the threshold is taken equal to about $500$~au, but this value is actually a function of eccentricity and inclination.
   
   As mentioned in Sect.~\ref{sec:adiff}, small bodies for which the semi-major axis undergoes a chaotic diffusion due to planetary scattering are often captured in mean-motion resonances with Neptune. Most of the time, these resonance crossings are only temporary. However, in some cases (that we will detail below), the resonant dynamics itself extracts the small body from the diffusive region by increasing its perihelion distance. In this case, the orbital elements have a very smooth, quasi-integrable, long-term evolution that can be studied by analytical (or semi-analytical) means. We will recall here the method proposed by \cite{HENRARD_1990,HENRARD_1993} and applied to the trans-Neptunian region by \cite{SAILLENFEST-etal_2016,SAILLENFEST-etal_2017a} and \cite{SAILLENFEST-LARI_2017}. We will also discuss its variant introduced independently by \cite{WISDOM_1985} and used for instance by \cite{SIDORENKO_2006,SIDORENKO_2018}. Section~\ref{sec:rescoo} presents the change of coordinates used to isolate the resonant angle and compute a semi-averaged Hamiltonian. Then, Sects.~\ref{sec:semisec} and \ref{sec:ressec} show how the resonant dynamics over intermediate and long timescales can be studied, and we summarise the implications of such a dynamics for trans-Neptunian objects.
   
   \subsection{The resonant coordinates}\label{sec:rescoo}
   Let us consider a resonant angle of the form
   \begin{equation}\label{eq:sigma}
      \sigma = k\lambda - k_p\lambda_p - (k-k_p)\varpi\,,
   \end{equation}
   where $k,k_p\in\mathbb{N}$ and $k>k_p$. In this expression, $\lambda$ and $\lambda_p$ are the mean longitudes of the small body and of the planet $p$ involved, and $\varpi=\omega+\Omega$. If this angle oscillates (librates), this means that the small body performs approximately $k_p$ orbits during $k$ orbits of the planet $p$. If the perturbation Hamiltonian $\varepsilon_\mathrm{P}\mathcal{H}_\mathrm{P}$ in Eq.~\eqref{eq:H0H1} is expanded in series of the eccentricity and of the inclination, we can show that it contains an infinite number of terms involving the angle $\sigma$, accompanied by combinations of $\varpi$ and $\Omega$ (see e.g. \citealp{MURRAY-DERMOTT_1999}). However, the semi-analytical method described below allows us to study all these terms at once, through the evolution of the only angle $\sigma$ from Eq.~\eqref{eq:sigma}. The inclusion of all terms is particularly important for highly eccentric and/or highly inclined orbits, such as the ones observed in the trans-Neptunian region. For such extreme orbits, numerous terms play a role simultaneously, giving rise to complex resonance structures that have little to do with the traditional ``resonance order'' $k-k_p$ (\citealp{PAN-SARI_2004,GALLARDO_2019}).
   
   Starting from the Delaunay elements (see Eq.~\ref{eq:Del}), we introduce the resonant canonical coordinates:
   \begin{equation}\label{eq:rescoo}
      \begin{pmatrix}
         \sigma \\
         \gamma \\
         u \\
         v
      \end{pmatrix}
      =
      \begin{pmatrix}
         k & -k_p & k_p & k_p \\
         c & -c_p & c_p & c_p \\
         0 & 0 & 1 & 0 \\
         0 & 0 & 0 & 1
      \end{pmatrix}
      \begin{pmatrix}
         \ell \\
         \lambda_p \\
         g \\
         h
      \end{pmatrix}
      \hspace{0.5cm};\hspace{0.5cm}
      \begin{pmatrix}
         \Sigma \\
         \Gamma \\
         U \\
         V
      \end{pmatrix}
      =
      \begin{pmatrix}
           -c_p & -c & 0 & 0 \\
            k_p &  k & 0 & 0 \\
           0 & 1 & 1 & 0 \\
           0 & 1 & 0 & 1
      \end{pmatrix}
      \begin{pmatrix}
         L \\
         \Lambda_p \\
         G \\
         H
      \end{pmatrix}\,,
   \end{equation}
   where $c$ and $c_p$ are integers chosen such that $c\,k_p-c_p\,k = 1$ (see \citealp{MILANI-BACCILI_1998}). If we assume that the particle is close or inside the resonance  considered, $\gamma$ and $\{\lambda_{i\neq p}\}$ are fast angles (orbital timescale $\propto 1$), $\sigma$ is a semi-slow angle (semi-secular timescale $\propto 1/\sqrt{\varepsilon_\mathrm{P}}$), and $(u,v)$ are slow angles (secular timescale $\propto 1/\varepsilon_\mathrm{P}$).
   
   As in Sect.~\ref{sec:sec}, we can now get rid of the fast angles by a near-identity change of coordinates. At first order to $\varepsilon_\mathrm{P}$, the ``semi-secular coordinates'' are obtained by averaging the Hamiltonian function over fast angles. In the semi-secular coordinates (that we write with the same symbols), the momenta $\Gamma$ and $\{\Lambda_{i\neq p}\}$ are constants of motion with arbitrary values. In particular, $\Gamma$ is conveniently chosen equal to zero, such that the coordinates become
   \begin{equation}\label{eq:SUV}
      \left\{
      \begin{aligned}
         \Sigma &= \frac{\sqrt{\mu a}}{k} \\\
         U &= \sqrt{\mu a}\left(\sqrt{1-e^2}-\frac{k_p}{k}\right) \\
         V &=  \sqrt{\mu a}\left(\sqrt{1-e^2}\cos I-\frac{k_p}{k}\right)
      \end{aligned}
      \right.
      \text{and}\hspace{0.3cm}
      \left\{
      \begin{aligned}
         \sigma &= k\lambda - k_p\lambda_p - (k-k_p)\varpi\\
         u &= \omega \\
         v &= \Omega\,.
      \end{aligned}
      \right.
   \end{equation}
   Thanks to the rotational symmetry induced by the circular and coplanar orbits of the planets, the semi-secular Hamiltonian does not depend on $v$, so that $V$ is a constant of motion. We are left with the two degrees of freedom $(\Sigma,\sigma)$ and $(U,u)$. Dropping unnecessary constants, the semi-secular Hamiltonian is
   \begin{equation}\label{eq:Kham}
      \mathcal{K} = \mathcal{K}_0(\Sigma) + \varepsilon_\mathrm{P}\mathcal{K}_\mathrm{P}(\Sigma,U,V,\sigma,u) + \mathcal{O}(\varepsilon_\mathrm{P}^2) \,,
   \end{equation}
   with
   \begin{equation}\label{eq:Kdet}
      \left\{
      \begin{aligned}
         \mathcal{K}_0 &= -\frac{\mu^2}{2(k\Sigma)^2} - n_pk_p\Sigma \,,\\
         \varepsilon_\mathrm{P}\mathcal{K}_\mathrm{P} &= - \sum_{i\neq p}\frac{\mu_i}{4\pi^2}\int_{0}^{2\pi}\!\!\!\int_{0}^{2\pi}\frac{1}{\|\mathbf{r}-\mathbf{r}_i\|}\,\mathrm{d}\ell\,\mathrm{d}\lambda_i \\
         &\ \ \ - \frac{\mu_p}{2\pi}\int_{0}^{2\pi}\left(\frac{1}{\|\mathbf{r}-\mathbf{r}_p\|}-\mathbf{r}\cdot\frac{\mathbf{r}_p}{\|\mathbf{r}_p\|^3}\right)\mathrm{d}\gamma \,.
      \end{aligned}
      \right.
   \end{equation}
   The required averages can be computed numerically, so that the formulas remain valid for any value of the orbital elements of the small body. This kind of semi-analytical procedure is now widely used for resonant problems in celestial mechanics involving high eccentricities and/or high inclinations (see e.g. \citealp{MILANI-BACCILI_1998,GALLARDO_2006b,GALLARDO_2006a,GALLARDO_2019,SIDORENKO_2006,SIDORENKO_2018,SAILLENFEST-etal_2016,SAILLENFEST-LARI_2017,PICHIERRI-etal_2017,BATYGIN-MORBIDELLI_2017}).
   
   We note that the evolution of the pair $(U,u)$ is secular by nature (frequency $\propto\varepsilon_\mathrm{P}$), whereas the evolution of the pair $(\Sigma,\sigma)$ is semi-secular (frequency $\propto\sqrt{\varepsilon_\mathrm{P}}$). The adiabatic approximation introduced by \cite{WISDOM_1985} and \cite{HENRARD_1993} consists in first studying the evolution of $(\Sigma,\sigma)$ for fixed values of $(U,u)$, and then using a new near-identity transformation in order to remove the semi-fast angle $\sigma$ from the Hamiltonian, similarly to what we did for obtaining Eq.~\eqref{eq:Kham}. However, this time, since the two characteristic frequencies are only separated by a factor $\sqrt{\varepsilon_\mathrm{P}}$, the neglected terms are of order $\varepsilon_\mathrm{P}^{3/2}$ (instead of $\varepsilon_\mathrm{P}^2$), meaning that this method is only accurate for very small values of~$\varepsilon_\mathrm{P}$. This is why it works particularly well for distant resonant trans-Neptunian objects, which undergo only very small perturbations from the planets.
   
   \subsection{Semi-secular evolution of $(\Sigma,\sigma)$}\label{sec:semisec}
\begin{figure}
      \centering
      \includegraphics[width=\textwidth]{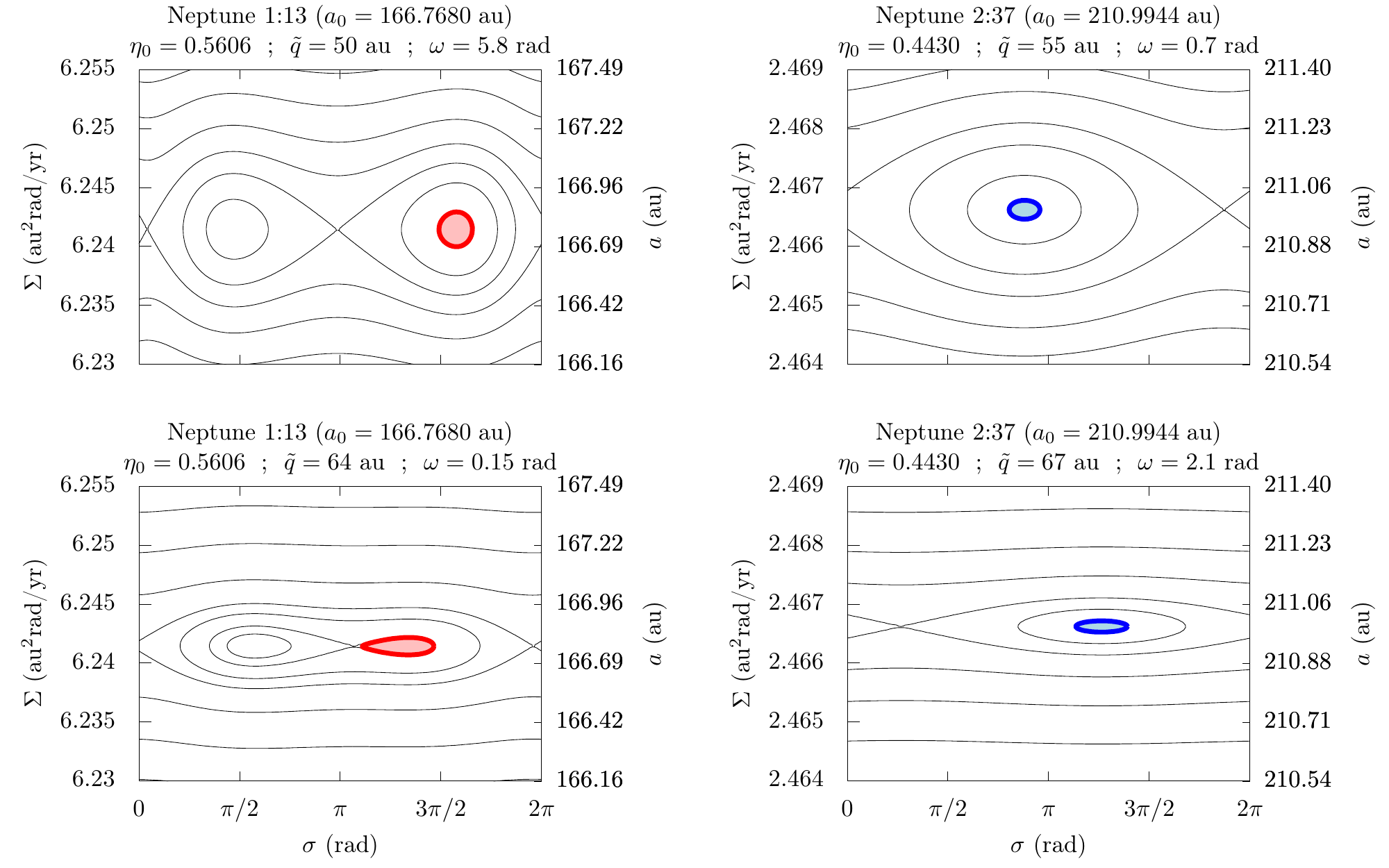}
      \caption{Level curves of the semi-secular Hamiltonian $\mathcal{K}$ in the plane $(\Sigma,\sigma)$. Different values of the parameters are used, indicated in title. The semi-major axis $a$ of the small body is given on the right; it is related to $\Sigma$ through Eq.~\eqref{eq:SUV}. The coloured curves are examples of resonant trajectories, and the areas $2\pi J$ enclosed are shown with lighter colours (see Eq.~\ref{eq:2piJ}). The resonances and parameters chosen are the same as Figs.~\ref{fig:Hressec1} and \ref{fig:Hressec2}, showing the correspondence between semi-secular and secular evolutions (top panels correspond to a time $t_1$ of the secular evolution, bottom panels correspond to a time $t_2>t_1$).}
      \label{fig:exres}
   \end{figure}

   Neglecting $\mathcal{O}(\varepsilon_\mathrm{P}^2)$, the dynamics driven by Hamiltonian~\eqref{eq:Kham} with $(U,u)$ fixed has only one degree of freedom. It is therefore integrable, and every possible trajectory corresponds to a specific level curve of $\mathcal{K}$ in the $(\Sigma,\sigma)$ plane. In the low-eccentricity low-inclination regime, $\mathcal{K}$ is very close to the pendulum Hamiltonian. This is not the case anymore for large eccentricities and/or large inclinations like the ones reached by numerous trans-Neptunian objects. Figure~\ref{fig:exres} shows some examples of the geometry of resonances in the trans-Neptunian region. For easier interpretation, the variable $U$ is replaced by the ``reference perihelion distance'' $\tilde{q}$ and the ``reference inclination'' $\tilde{I}$ introduced by \cite{SAILLENFEST-etal_2016}. They correspond to the actual $q$ and $I$ of the small body whenever its semi-major axis is equal to $a_0$, where $a_0$ is a wisely-chosen constant (average centre of the resonance). Likewise, the parameter $V$ is replaced by
   \begin{equation}
      \eta_0 = \frac{V}{\sqrt{\mu a_0}} + \frac{k_p}{k} = \sqrt{1-\tilde{e}^2}\cos\tilde{I} \,,
   \end{equation}
   where $\tilde{q} = a_0(1-\tilde{e})$. Using diagrams like those shown in Fig.~\ref{fig:exres}, one can compute numerically various properties of the resonances, including the exact value of their width for any eccentricity and inclination.
   
   As shown by \cite{GALLARDO_2006b} and \cite{SAILLENFEST-etal_2016}, an inner double island (left panels of Fig.~\ref{fig:exres}) is always present for resonances with $k_p=1$, provided that the eccentricity is high enough. Such a structure had already been noted by \cite{SCHUBART_1964,BEAUGE_1994,MORBIDELLI-etal_1995}. This phenomenon can be explained qualitatively by looking at the expansion of $\varepsilon_\mathrm{P}\mathcal{K}_\mathrm{P}$ in series of the eccentricity. Indeed, using the explicit formulas by \cite{ELLIS-MURRAY_2000}, we see that resonances with $k_p=1$ are the only ones that feature a contribution in the indirect part $\mathcal{R}_\mathrm{I}$ of the perturbation. For a given interval of eccentricity, this contribution partially cancels the lowest-order term of the direct part $\mathcal{R}_\mathrm{D}$, meaning that the dominant cosine term is not $\cos\sigma$ anymore, but $\cos(2\sigma)$. This explains the presence of a double island of resonance in the plane $(\Sigma,\sigma)$, with an internal separatrix. \cite{PAN-SARI_2004} obtain similar findings in the planar case using high-eccentricity mapping techniques.
   
   \subsection{Secular evolution of $(U,u)$}\label{sec:ressec}
   In order to remove the semi-fast component of the dynamics from the Hamiltonian $\mathcal{K}$, we need to replace $(\Sigma,\sigma)$ by action-angle coordinates. Such coordinates are composed of one angle $\theta$ evolving linearly with time, and one constant momentum $J$. As recalled by \cite{HENRARD_1990}, this constant momentum can be written
   \begin{equation}\label{eq:2piJ}
      2\pi J = \frac{1}{2}\oint\left(\Sigma\,\mathrm{d}\sigma-\sigma\,\mathrm{d}\Sigma\right) \,,
   \end{equation}
   which corresponds to the signed area enclosed or stretched by the trajectory in the plane $(\Sigma,\sigma)$. The constant $J$ is said to be the adiabatic invariant of the system. By definition of the action-angle coordinates, the Hamiltonian does not depend on $\theta$. This leads to the definition of the secular Hamiltonian function, which can be formally written\footnote{In \cite{SAILLENFEST-etal_2016}, it is written that terms of order $\xi=\varepsilon_\mathrm{P}^{1/2}$ are neglected, instead of $\varepsilon_\mathrm{P}^{3/2}$. Indeed, their Hamiltonian was implicitly divided by $\varepsilon_\mathrm{P}$.}
   \begin{equation}\label{eq:Fressec}
      \mathcal{F} = \mathcal{F}_0(J,U,V,u) + \mathcal{O}\left(\varepsilon_\mathrm{P}^{3/2}\right) \,.
   \end{equation}
   The adiabatic invariant $J$ becomes a parameter of the model, along with the momentum $V$. The secular system has thus a single degree of freedom, and all the possible trajectories can be represented by the level curves of $\mathcal{F}$ in the plane $(U,u)$. It should be noted that contrary to Sect.~\ref{sec:sec}, the model now includes the effects of a mean-motion resonance. The term ``secular'' is thus somehow improper here. Whenever there is an ambiguity, we refer to this model as being a ``resonant secular'' theory, not to be confused with the ``secular'' theory from Sect.~\ref{sec:sec}.
   
\begin{figure}
      \centering
      \includegraphics[width=0.7\textwidth]{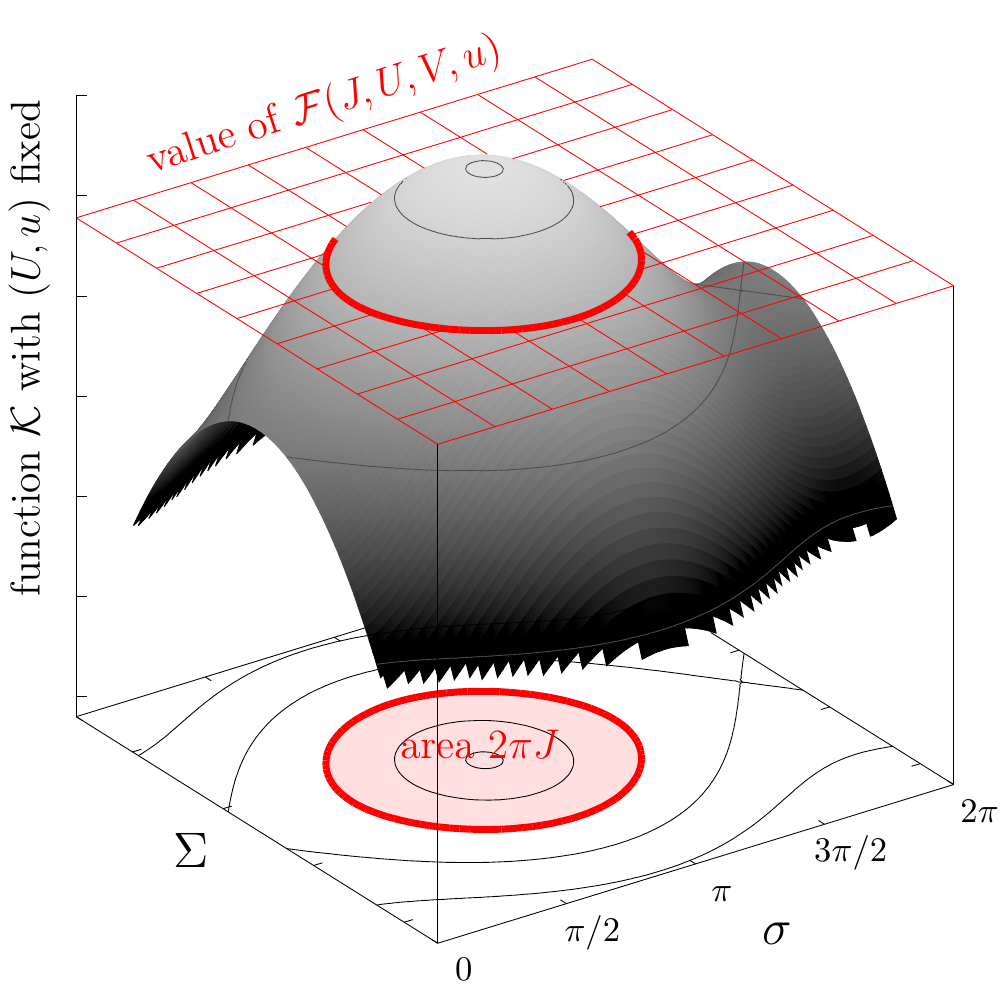}
      \caption{Illustration of the link between the adiabatic invariant $J$ and the value of the resonant secular Hamiltonian $\mathcal{F}$. The trajectory of the slow variables $(U,u)$ is such that both $J$ and $\mathcal{F}$ are constant while the three-dimensional shape is deformed. Henrard's method consists in looking for $\mathcal{F}$ knowing $J$. Wisdom's method consists in looking for $J$ knowing $\mathcal{F}$.}
      \label{fig:comp}
   \end{figure}

   The method detailed by \cite{HENRARD_1993} consists in computing numerically the secular Hamiltonian $\mathcal{F}$ and its partial derivatives. For a given value of $J$, the computation of the secular Hamiltonian at one point $(U,u)$ requires to look numerically in the plane $(\Sigma,\sigma)$ for the level curve of $\mathcal{K}$ that encloses or stretches an area $2\pi J$. The search is quite simple for $J=0$ (zero-amplitude oscillations of $\Sigma$ and $\sigma$) because we just need to pick up the value of $\mathcal{K}$ at its maximum. This is why many authors, like \cite{BEUST_2016} or \cite{PICHIERRI-etal_2017}, limit their analyses to the case $J=0$. For larger-amplitude oscillations or circulation of $\sigma$, we need to apply a Newton algorithm to Eq.~\eqref{eq:2piJ} as a function of the initial position $(\Sigma_i,\sigma_i)$. Once an initial position producing the required value of $J$ has been found, the secular Hamiltonian is simply
   \begin{equation}
      \mathcal{F}_0(J,U,V,u) = \mathcal{K}(\Sigma_i,U,V,\sigma_i,u) \,.
   \end{equation}
   No further averaging is needed. Figure~\ref{fig:comp} illustrates the graphical meaning of $J$ and the corresponding value of $\mathcal{F}$. Whenever the secular evolution of $(U,u)$ drives the semi-secular variables $(\Sigma,\sigma)$ through a separatrix (see Fig.~\ref{fig:exres}, bottom left panel), the adiabatic approximation breaks down. This means that $J$ can make unpredictable jumps (\citealp{HENRARD-MORBIDELLI_1993} speak of a ``stochastic layer''), and that a new phase portrait is required after the crossing.
   
   The variant introduced by \cite{WISDOM_1985} consists in using the value of $\mathcal{F}$ as parameter, and representing the solutions of the dynamics as the level curves of the adiabatic invariant $J$ in the plane $(U,u)$. The roles of $\mathcal{F}$ and $J$ are thus inverted with respect to Henrard's method. This inversion makes Wisdom's method easier to implement than Henrard's. It also removes the need to compute a new phase portrait whenever a separatrix crossing is encountered (see e.g. \citealp{WISDOM_1985,SIDLICHOVSKY_2005,SIDORENKO_2006,SIDORENKO-etal_2014,SIDORENKO_2018}). The two methods are however strictly equivalent and they describe the same solutions.
   
   The resonant secular theory outlined here is very efficient for characterising the long-term dynamics of distant trans-Neptunian objects trapped in mean-motion resonance with Neptune. It has be extensively used by \cite{SAILLENFEST-etal_2017a} to explore the distant trans-Neptunian region, showing pathways to high perihelion distances, as well as a ``trapping mechanism'' able to maintain the objects on very distant orbits for billions of years. This trapping mechanism is associated with the crossing of the innermost separatrix of resonances of type $1:k$ (see the bottom left panel of Fig.~\ref{fig:exres}), forcing the trajectory to switch from asymmetric librations to horseshoe-type librations. An example of such an evolution is given below. The application of the resonant secular model to the known resonant objects is also very informative, since it shows graphically which observed orbits require a complex scenario (as the planetary migration or an external perturber), and which ones can be explained by the influence of the known planets in their current state \citep{SAILLENFEST-LARI_2017}. In the latter case, the dynamical history of small bodies can be tracked back to their capture in resonance.
   
\begin{figure}
      \centering
      \includegraphics[width=\textwidth]{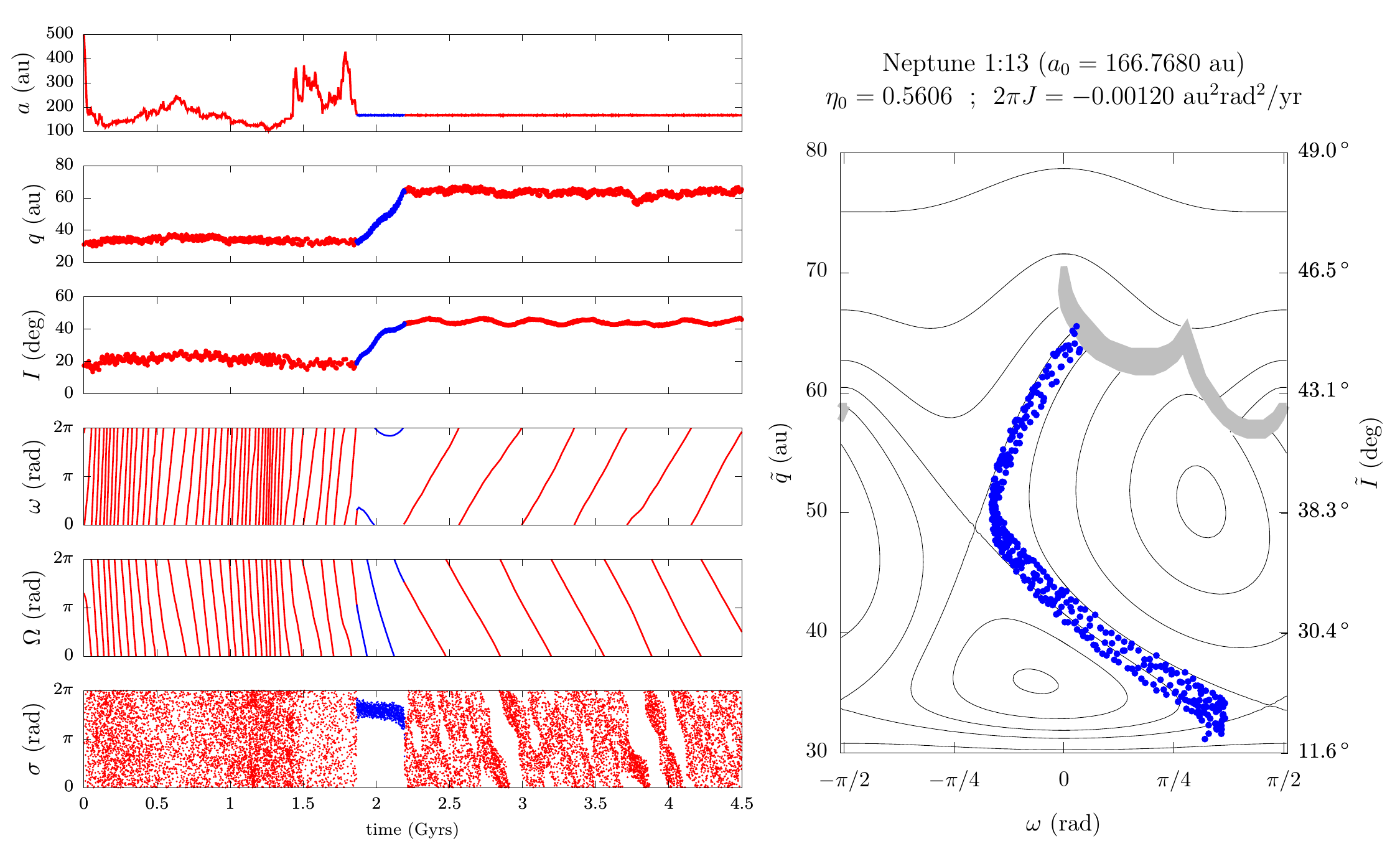}
      \caption{Comparison between numerical integration and the resonant secular semi-analytical model, adapted from \cite{SAILLENFEST-etal_2017a}. The left panel shows the evolution of the non-averaged orbital elements of a fictitious trans-Neptunian object perturbed by the four giant planets of the solar system. In the right panel, the blue portion of the numerical trajectory is plotted against the level curves of the secular Hamiltonian function $\mathcal{F}$ (see Eq.~\ref{eq:Fressec}), with the parameters given in title. The inclination values on the right are obtained from the constancy of $\eta_0$. The grey region is forbidden for these values of the parameters. Crossing the grey region means that the coordinates $(\Sigma,\sigma)$ cross a separatrix (see Fig.~\ref{fig:exres}, bottom left panel). This triggers a different kind of dynamics, characterised in this example by an almost constant value of $q$ (see left panel).}
      \label{fig:Hressec1}
   \end{figure}

\begin{figure}
      \centering
      \includegraphics[width=\textwidth]{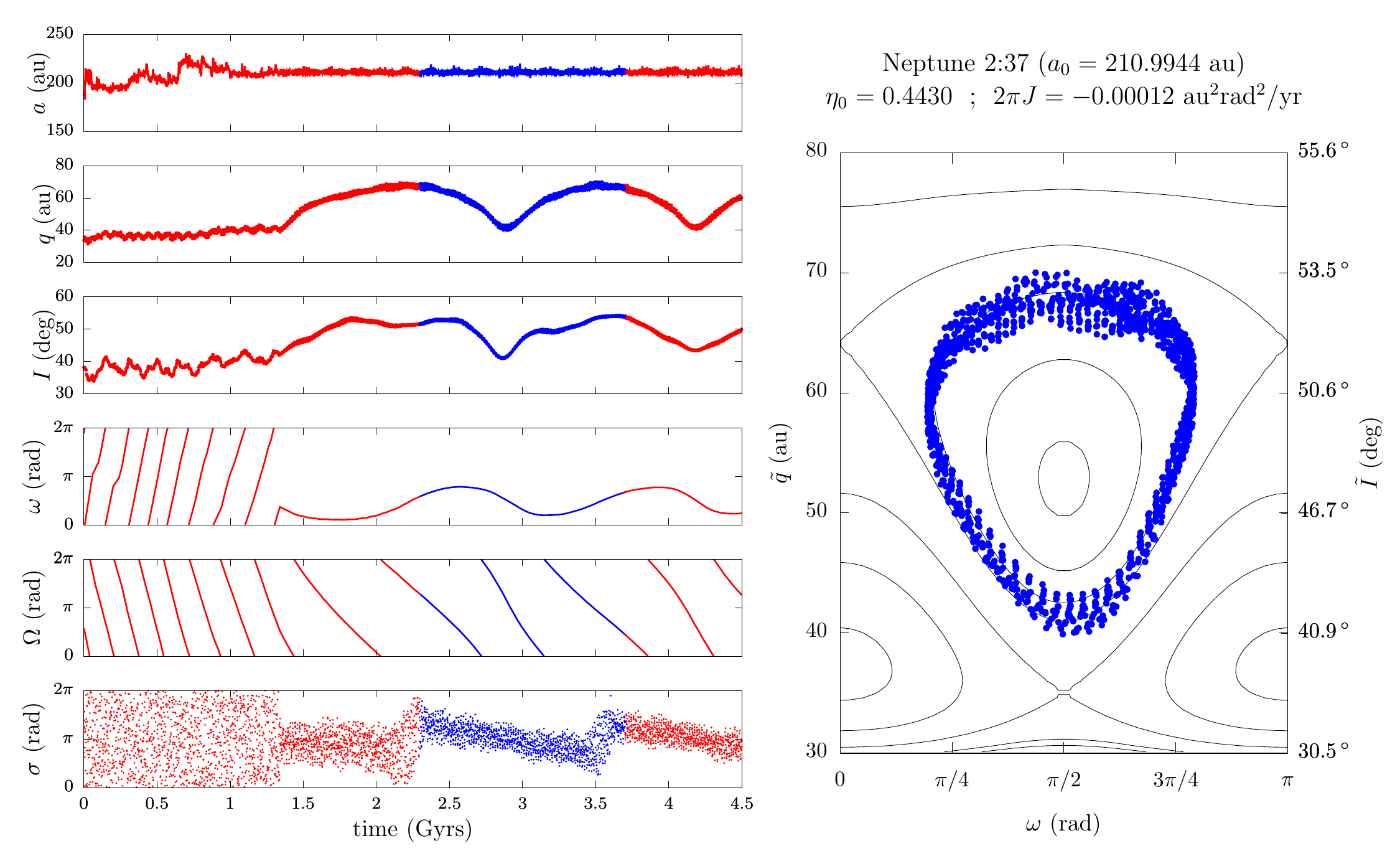}
      \caption{Same as Fig.~\ref{fig:Hressec1} for another fictitious particle, trapped in another high-order mean-motion resonance.}
      \label{fig:Hressec2}
   \end{figure}

   Figures~\ref{fig:Hressec1} and \ref{fig:Hressec2} compare non-averaged numerical integrations to the results given by the semi-analytical model. Both figures show the behaviour of a fictitious trans-Neptunian object initially located in the diffusion region (bottom part of Fig.~\ref{fig:zones}, see Sect.~\ref{sec:adiff}). Then, at some point, the small body encounters a high-order mean-motion resonance with Neptune ($1:13$ and $2:37$, respectively). As predicted by the semi-analytical model, the resonant link with Neptune is such that the long-term dynamics removes the small body from the diffusive region. It reaches therefore the region labelled ``isolated planetary resonances'' in Fig.~\ref{fig:zones}. \cite{SAILLENFEST-etal_2017a} even found resonant pathways towards perihelion distances larger than $100$~au (but with very low-probability entrance for actual small bodies). Once arrived in this region, the two scenarios from Figs.~\ref{fig:Hressec1} and \ref{fig:Hressec2} differ. In Fig.~\ref{fig:Hressec1}, the trajectory in the plane $(\Sigma,\sigma)$ crosses a separatrix, triggering a protective mechanism that prevents the small body from going back to the diffusive region for billions of years. Even without such a mechanism, the resonant link with Neptune weakens as the perihelion distance grow, making the small body vulnerable to any other perturbation such as the small but non-zero eccentricities and inclinations of the giant planets. This drop-off in resonance strength can eventually leave small bodies slightly out of resonance at high perihelion distances \citep{SAILLENFEST-etal_2017a,GOMES-etal_2005}. In Fig.~\ref{fig:Hressec2}, on the contrary, the trajectory cycles back down towards the diffusive region, where the resonant link could be broken again. These cycles in and out of the diffusive region have also been at play during Neptune's orbital migration, leading to the formation of detached non-resonant bodies (see e.g. \citealp{GOMES_2003,GOMES-etal_2005,GOMES_2011}). This mechanism can be used to constrain the properties of Neptune's orbital migration in the late stages of the formation of the solar system \citep{NESVORNY-VOKROUHLICKY_2016,LAWLER-etal_2019}.
   
\begin{figure}
      \centering
      \includegraphics[width=\textwidth]{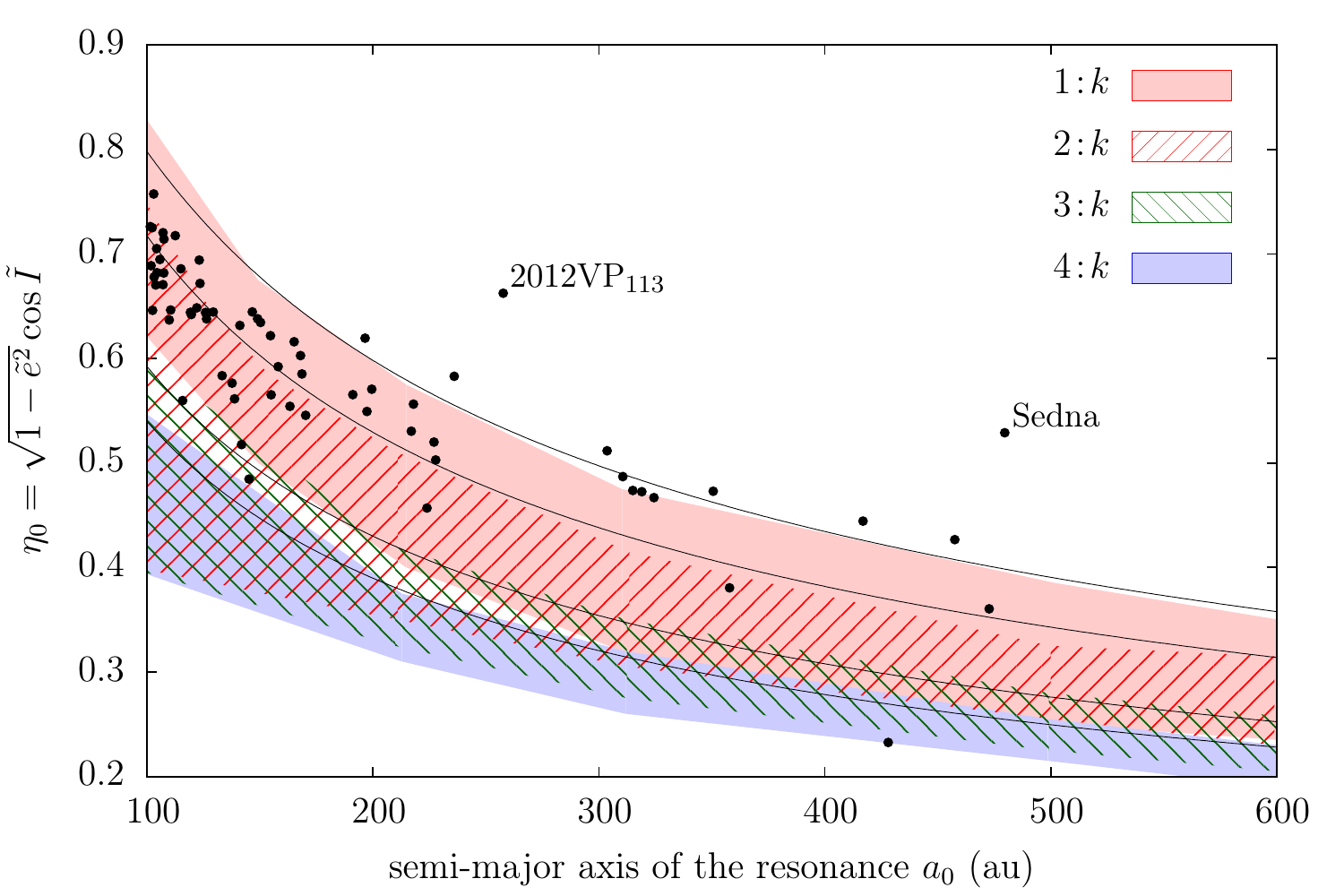}
      \caption{Interval of parameters producing libration zones of $(\omega,\tilde{q})$ for various types of mean-motion resonances $k_p:k$ with Neptune (adapted from \citealp{SAILLENFEST-etal_2017a}). The figure is calculated for $J=0$ and the plot is restricted to prograde orbits ($\eta_0>0$). In order to guide the eye, black curves show the value of $\eta_0$ for $\tilde{I}=0$ and constant $\tilde{q}$ (from top to bottom, $\tilde{q} = 40$, $30$, $19$ and $16$~au). Increasing $\tilde{I}$ would shift the curves down. The black spots represent the position of the currently known small bodies with $a>100$~au and $q>30$~au according to the JPL Small Body Database (\texttt{https://ssd.jpl.nasa.gov/}), using $a$ for $a_0$ and $\sqrt{1-e^2}\cos I$ for $\eta_0$.}
      \label{fig:interet}
   \end{figure}

   Only a small fraction of captures in mean-motion resonance lead to strong orbital changes like those illustrated in Figs.~\ref{fig:Hressec1} and \ref{fig:Hressec2}, with a change of dynamical regime. The ability of mean-motion resonances with Neptune to strongly modify the orbits of small bodies was measured by \cite{SAILLENFEST-etal_2017a} in the space of orbital parameters. First, the resonance must not be too distant ($a\lesssim 500$~au, see Fig.~\ref{fig:zones}). Second, the capture must be deep enough, that is, with a parameter $J$ close to zero, corresponding to small-amplitude oscillations of the resonant angle (in the schematic picture of Fig.~\ref{fig:tubes}, such captures can only occur at the bottom extremity of the resonant tubes, where they are narrower). Finally, the parameter $\eta_0$ must lie in a specific range shown in Fig.~\ref{fig:interet}, dubbed ``range of interest''. Since the condition from Fig.~\ref{fig:interet} combines both $I$ and $e$, we stress that an initial high inclination is not necessary to trigger large variations of perihelion distance and inclination. Indeed, the curve that corresponds to $\tilde{I}=0$ and $\tilde{q}=30$~au lie right in the middle of the range of interest for resonances of type $1:k$ (see the black curves). For other types of resonances, however, small bodies cannot reach the required range with $\tilde{I}=0$ and $\tilde{q}$ beyond Neptune. They either need a larger inclination at the time of the resonance capture, or a perihelion distance in the region of Uranus. Interestingly, Fig.~\ref{fig:interet} shows that most of the observed trans-Neptunian objects would fall in some range of interest (especially for resonances of type $1:k$) in case of capture in mean-motion resonance. Hence, their orbits can be readily explained by a combination of scattering and resonance trapping (and the late migration of Neptune is accounted for their eventual release out of resonance, see Sect.~\ref{sec:sculpt}). The two most notable outliers, $2012$VP$_{113}$ and (90377) Sedna, are explicitly labelled in Fig.~\ref{fig:interet}: no mean-motion resonance with Neptune could possibly have shaped their orbits. The same conclusions were obtained numerically by \cite{LAWLER-etal_2019}.
   
\section{Galactic tides}\label{sec:gt}
   For large semi-major axes (say $a>1600$~au at least) and perihelion distances well separated from the orbit of Neptune, trans-Neptunian objects are quite insensible to the detailed structure of the planetary region. They simply orbit around the barycentre of the solar system in an almost unperturbed two-body problem. For such distant objects, however, external forces, like the gravitational perturbations from the galactic tides, are noticeable (upper right portion of Fig.~\ref{fig:zones}). Hence, in this region of orbital elements, the planetary perturbations can be neglected and the galactic tides can be studied as a perturbation to the barycentric trajectory. The Hamiltonian function describing the orbital dynamics of a small body is therefore similar to Eq.~\eqref{eq:Hfirst}, but with a different perturbation:
   \begin{equation}
      \mathcal{H} = \mathcal{H}_0(L) + \varepsilon_\mathrm{G}\mathcal{H}_\mathrm{G}(L,G,H,\ell,g,h,t) \,,
   \end{equation}
   where G stands for ``Galaxy''.
   
   We consider the coordinates $(X,Y,Z)$ of the small body in a fixed reference frame centred on the Sun, where the $(X,Y)$ plane is the galactic plane. We note $(X',Y',Z')$ the coordinates of the small body in an analogous reference frame, but for which at any time the $X'$ axis points towards the galactic centre. Because of the motion of the Sun in the Galaxy, the latter reference frame is rotating. At lowest-order of approximation, the Sun describes a circular orbit with constant velocity lying in the galactic plane (see e.g. \citealp{FOUCHARD_2004}). We have in this case the relation
   \begin{equation}
      \begin{pmatrix}
         X' \\
         Y' \\
         Z'
      \end{pmatrix}
      =
      \begin{pmatrix}
          \cos\theta &  \sin\theta & 0 \\
         -\sin\theta &  \cos\theta & 0 \\
                   0 &           0 & 1
      \end{pmatrix}
      \begin{pmatrix}
         X \\
         Y \\
         Z
      \end{pmatrix}
      \,,
   \end{equation}
   where the time derivative of $\theta$ is a constant and corresponds to the angular velocity of the galactic centre seen from the Sun. In the following, we write it $\nu_\mathrm{G}$. In the quadrupolar approximation, the Hamiltonian function describing the orbital perturbation of the small body from the galactic tides can be written
   \begin{equation}
      \varepsilon_\mathrm{G}\mathcal{H}_\mathrm{G} = \nu_\mathrm{G}P_\theta + \mathcal{G}_1\frac{X'^2}{2} + \mathcal{G}_2\frac{Y'^2}{2} + \mathcal{G}_3\frac{Z'^2}{2} \,,
   \end{equation}
   where the momentum $P_\theta$ is conjugate to the angle $\theta$; it has been introduced such that the Hamiltonian function is autonomous. $\mathcal{G}_1$, $\mathcal{G}_2$, and $\mathcal{G}_3$ are constants encompassing the shape of the Galaxy, its mass density, and the inertial forces due to the rotation of the frame. See for instance \cite{VOKROUHLICKY-etal_2019} for a discussion about the values of $\mathcal{G}_1$, $\mathcal{G}_2$, and $\mathcal{G}_3$. They are considered constant in first approximation, even though they actually depend on the precise location of the Sun in the galaxy \citep{KAIB-etal_2011}. Here we will stick to the approximation $\mathcal{G}_2 = -\mathcal{G}_1$, for which we obtain
   \begin{equation}
      \varepsilon_\mathrm{G}\mathcal{H}_\mathrm{G} = \nu_\mathrm{G}P_\theta + \varepsilon_{\mathrm{G}_\mathrm{V}}\mathcal{H}_{\mathrm{G}_\mathrm{V}} + \varepsilon_{\mathrm{G}_\mathrm{R}}\mathcal{H}_{\mathrm{G}_\mathrm{R}} \,,
   \end{equation}
   where
   \begin{equation}
      \left\{
      \begin{aligned}
         \varepsilon_{\mathrm{G}_\mathrm{V}}\mathcal{H}_{\mathrm{G}_\mathrm{V}} &= \mathcal{G}_3\frac{Z^2}{2} \\
         \varepsilon_{\mathrm{G}_\mathrm{R}}\mathcal{H}_{\mathrm{G}_\mathrm{R}} &= \mathcal{G}_2\left(\frac{Y^2-X^2}{2}\cos(2\theta)-XY\sin(2\theta)\right) \,.
      \end{aligned}
      \right.
   \end{equation}
   The symbols V and R are used here in reference to the vertical and radial components of the galactic tides, respectively.
   
   The perturbation $\varepsilon_\mathrm{G}\mathcal{H}_\mathrm{G}$ being very small with respect to the Keplerian part $\mathcal{H}_0$, it acts on a much longer timescale. Therefore, we can use a perturbative approach to order one, similar to what we did in Sect.~\ref{sec:sec}. The resulting Hamiltonian function is obtained by averaging $\varepsilon_\mathrm{G}\mathcal{H}_\mathrm{G}$ over an orbital period of the unperturbed Keplerian orbit. Neglecting $\mathcal{O}(\varepsilon_\mathrm{G}^2)$, the momentum conjugate to $\ell$ becomes a constant of motion, which implies the conservation of the secular semi-major axis (that we still denote $a$). Dropping the constant terms, the secular Hamiltonian is
   \begin{equation}\label{eq:Fgal}
      \mathcal{M} = \nu_\mathrm{G}P_\theta + \varepsilon_{\mathrm{G}_\mathrm{V}}\mathcal{M}_{\mathrm{G}_\mathrm{V}} + \varepsilon_{\mathrm{G}_\mathrm{R}}\mathcal{M}_{\mathrm{G}_\mathrm{R}} \,.
   \end{equation}
   Using the explicit notations
   \begin{equation}\label{eq:epsG}
      \varepsilon_{\mathrm{G}_\mathrm{V}} = a^2\mathcal{G}_3
      \hspace{0.5cm},\hspace{0.5cm}
      \varepsilon_{\mathrm{G}_\mathrm{R}} = a^2\mathcal{G}_2 \,,
   \end{equation}
   the two parts can be expressed as
   \begin{equation}\label{eq:FGvr}
      \left\{
      \begin{aligned}
         \mathcal{M}_{\mathrm{G}_\mathrm{V}} &= \frac{\sin^2I_\mathrm{G}}{4}\left(1+\frac{3}{2}e^2-\frac{5}{2}e^2\cos(2\omega_\mathrm{G})\right) \,,\\
         \mathcal{M}_{\mathrm{G}_\mathrm{R}} &= -\frac{1}{4}\left(1+\frac{3}{2}e^2\right)\cos(2\Omega_\mathrm{G}-2\theta)\sin^2I_\mathrm{G} \\
         &+ \frac{5}{4}e^2\Bigg(\sin(2\omega_\mathrm{G})\sin(2\Omega_\mathrm{G}-2\theta)\cos I_\mathrm{G} \\
         &\hspace{1cm}
         - \cos(2\omega_\mathrm{G})\cos(2\Omega_\mathrm{G}-2\theta)\frac{1+\cos^2I_\mathrm{G}}{2}\Bigg) \,,
      \end{aligned}
      \right.
   \end{equation}
   where $(a,e,I_\mathrm{G},\omega_\mathrm{G},\Omega_\mathrm{G})$ are the Keplerian elements of the small body expressed in the $(X,Y,Z)$ reference frame, that is, with the reference plane in the galactic plane.
   
   The dynamics driven by the secular Hamiltonian $\mathcal{M}$ in Eq.~\eqref{eq:Fgal} has been studied by many authors (see \citealp{FOUCHARD_2004,BREITER-etal_2008} and references therein). Extremely fast algorithms have been developed to compute its solutions \citep{BREITER-etal_2007,FOUCHARD-etal_2007}. Taking the values of $\mathcal{G}_2$ and $\mathcal{G}_3$ from the literature, however, one can notice that $\mathcal{G}_3$ is one order of magnitude larger than $\mathcal{G}_2$. The dynamics is therefore largely dominated by $\varepsilon_{\mathrm{G}_\mathrm{V}}\mathcal{M}_{\mathrm{G}_\mathrm{V}}$, and the radial component of the galactic tides only acts as a long-term modulation of the solutions, which remains small in a realistic amount of time. This is why, in order to draw a qualitative picture of the dynamics, authors often neglect $\varepsilon_{\mathrm{G}_\mathrm{R}}$ with respect to $\varepsilon_{\mathrm{G}_\mathrm{V}}$, or they average the effects of $\varepsilon_{\mathrm{G}_\mathrm{R}}\mathcal{M}_{\mathrm{G}_\mathrm{R}}$ (see e.g. \citealp{HEISLER-TREMAINE_1986,BRASSER_2001}). This has the enormous advantage of making the dynamics integrable, and the solutions can even be written with explicit analytical formulas involving elliptic integrals (see \citealp{BREITER-RATAJCZAK_2005}, \citealp{HIGUCHI-etal_2007}, \citealp{HIGUCHI-KOKUBO_2015}).
   
\begin{figure}
      \centering
      \includegraphics[width=\textwidth]{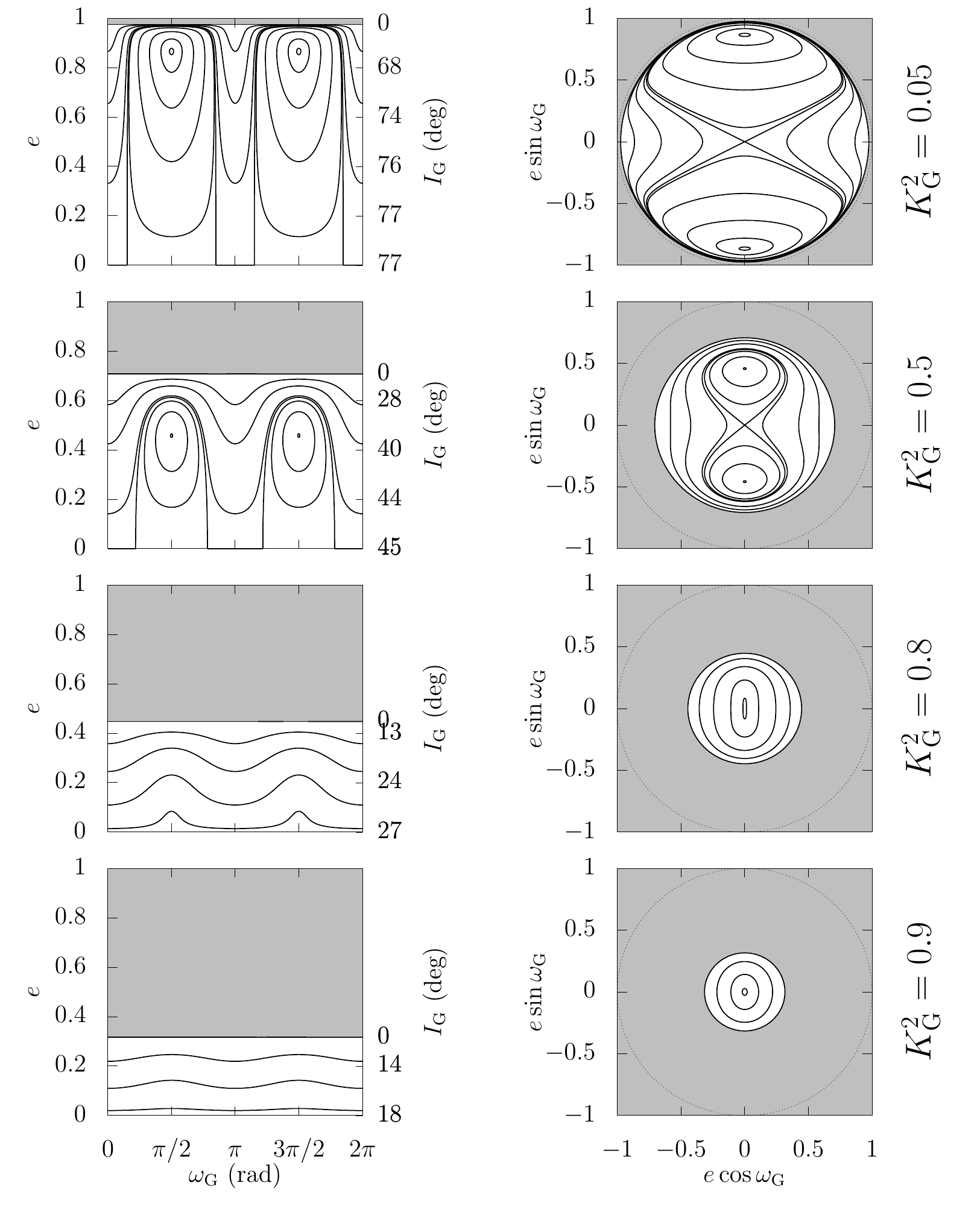}
      \caption{Level curves of the secular Hamiltonian $\mathcal{M}_{\mathrm{G}_\mathrm{V}}$, describing the long-term effect of galactic tides. Each row shows the trajectories allowed for a given value of $K_\mathrm{G}$, written on the right. The two columns show the same level curves in two different sets of coordinates. The grey regions are forbidden. On the left column, the right axis shows the orbital inclination obtained through the constancy of $K_\mathrm{G}$, assuming a prograde orbit.}
      \label{fig:Hgalniv}
   \end{figure}

   We recall here the main features of the dynamics driven by $\varepsilon_{\mathrm{G}_\mathrm{V}}\mathcal{M}_{\mathrm{G}_\mathrm{V}}$ taken alone. First of all, the Hamiltonian function does not depend on $\Omega_\mathrm{G}$. This means that its conjugate momentum is a constant of motion, in a similar way as in Sect.~\ref{sec:sec}. We introduce the ``galactic Kozai constant'':
   \begin{equation}
      K_\mathrm{G} = \sqrt{1-e^2}\cos I_\mathrm{G} \,,
   \end{equation}
   which can be used as a parameter of the Hamiltonian function. In turn, the system has only one degree of freedom, and every trajectory can be represented as a level curve of $\mathcal{M}_{\mathrm{G}_\mathrm{V}}$ in the $(\omega_\mathrm{G},e)$ plane. Figure~\ref{fig:Hgalniv} shows the level curves of $\mathcal{M}_{\mathrm{G}_\mathrm{V}}$ for different values of $K_\mathrm{G}$. The limit $I_\mathrm{G}=0$ or $180^\text{o}$ is a stable fixed point whatever the eccentricity. It results in a frozen orbit. Using $K_\mathrm{G}$ as parameter, this is equivalent to the condition
   \begin{equation}
      e^2 = 1-K_\mathrm{G}^2
      \ \iff\ 
      \cos^2I_\mathrm{G} = 1 \,.
   \end{equation}
   It corresponds to the border of the forbidden regions in Fig.~\ref{fig:Hgalniv}. The limit $e=0$ is a fixed point with circulating $\Omega_\mathrm{G}$, but it is unstable for $K_\mathrm{G}^2<4/5$, that is, for $27^\text{o}<I_\mathrm{G}<153^\text{o}$. For $K_\mathrm{G}^2<4/5$, there are two additional fixed points located at $\omega_\mathrm{G} = \pi/2$ and $3\pi/2$, with
   \begin{equation}\label{eq:eqpG}
      e^2 = 1 - \frac{\sqrt{5K_\mathrm{G}^2}}{2}
      \ \iff\ 
      e^2 = 1 - \frac{5}{4}\cos^2I_\mathrm{G} \,.
   \end{equation}
   These fixed points are stable, still with circulating $\Omega_\mathrm{G}$. As in Sect.~\ref{sec:sec}, the conservation of $K_\mathrm{G}$ implies that the orbit cannot become retrograde if it is prograde, and vice versa (but this time, this concerns the galactic inclination $I_\mathrm{G}$, not the ecliptic one $I$). Moreover, $\Omega_\mathrm{G}$ is always decreasing if $I_\mathrm{G}<90^\text{o}$ and always increasing if $I_\mathrm{G}>90^\text{o}$ (the period of its linear part can be found in \citealp{HIGUCHI-etal_2007}). Finally, as noted by \cite{HAMILTON-RAFIKOV_2019} about the same Hamiltonian in a different context, for $K_\mathrm{G}^2 < 4/5$ the eccentricity value at the stable equilibrium points (see Eq.~\ref{eq:eqpG}) is a lower bound of the maximum eccentricity reached by any trajectory.
   
\begin{figure}
      \centering
      \includegraphics[width=\textwidth]{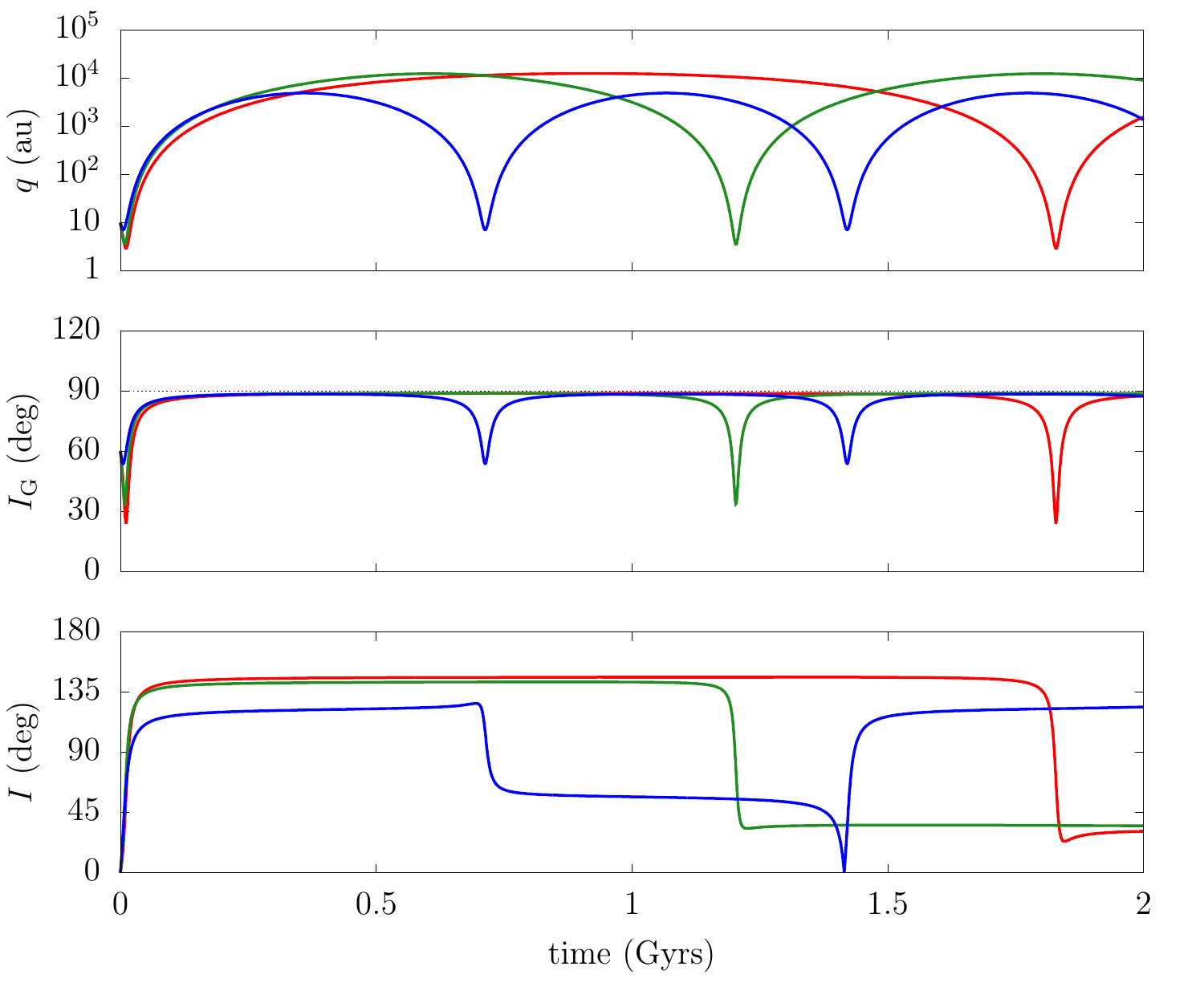}
      \caption{Examples of trajectories of long-period comets under the action of the galactic tides. The three trajectories represented have a semi-major axis of $20\,000$~au. At $t=0$, they start with $q=10$~au, in the ecliptic plane ($I=0^\text{o}$, i.e. $I_\mathrm{G}\approx 60^\text{o}$), and with different values of $\omega_\mathrm{G}$. Then, they evolve according to Hamiltonian $\varepsilon_{\mathrm{G}_\mathrm{V}}\mathcal{M}_{\mathrm{G}_\mathrm{V}}$.}
      \label{fig:cometsOC}
   \end{figure}

   Using the analytical expression of the solutions, \cite{HIGUCHI-etal_2007} showed that small bodies spend most of the time in the lower-eccentricity (i.e. higher-galactic-inclination) portion of their trajectory, and pass relatively quickly in the higher-eccentricity (i.e. lower-galactic-inclination) portion. Due to galactic tides, long-period comets residing in the Oort cloud undergo long-term oscillations of the perihelion distance that can bring them in and out of the planetary region. Their high-eccentricity passages through the planetary region are however very fast compared to the remaining portion of their orbital cycles. Figure~\ref{fig:cometsOC} shows examples of such trajectories. Due to the conservation of $K_\mathrm{G}$, the cycles of perihelion distance $q$ are correlated with cycles of the galactic inclination $I_\mathrm{G}$. The accumulation of orbits at high inclination $I_\mathrm{G}$, where they spend most of their time, is a typical feature of simulations (see e.g. \citealp{DYBCZYSKI-etal_2008,VOKROUHLICKY-etal_2019}). This can be easily understood by looking at the level curves of the Hamiltonian for very eccentric orbits, resulting in a small value of $K_\mathrm{G}$ (see Fig.~\ref{fig:Hgalniv}). Importantly, there is no such correlation between $q$ and the ecliptic inclination $I$, because $I$ depends on the value of $\Omega_\mathrm{G}$ (see Eq.~\ref{eq:cosI} below), which does not have the same period as $\omega_\mathrm{G}$ (see \citealp{HIGUCHI-etal_2007}). This means that from one low-perihelion-distance passage to the next one, the ecliptic inclination $I$ can change drastically. This partly explains why the Oort cloud appears isotropic. Moreover, passages at small perihelion distances can have dramatic consequences, since the small body enters the region of planetary scattering (see Fig.~\ref{fig:zones}). For instance, its semi-major axis can brutally decrease towards regions where the galactic tides are turned off. Oort cloud objects can therefore become scattered disc objects or centaurs with any inclination $I$ \citep{EMELYANENKO-etal_2007,KAIB-etal_2009,BRASSER-etal_2012,GOMES-etal_2015,KAIB-etal_2019}, that may further evolve into Halley-type comets \citep{LEVISON-etal_2001,NESVORNY-etal_2017}.
   
   The period of the orbital cycles induced by galactic tides strongly decreases with the semi-major axis of the small body. Authors generally define the ``outer Oort cloud'' as the region where galactic tides are able to move the perihelion of comets from outside $15$~au to inside $5$~au in less than one orbital period \citep{FOUCHARD_2010,KAIB-etal_2009}. Such comets, which have $a\gtrsim 20\,000$~au, avoid a potential ejection by Saturn or Jupiter and become directly observable from Earth. Comets from the ``inner Oort cloud'', having $a\lesssim 20\,000$~au, pass many times at perihelion during their high-eccentricity phase, which, for small-enough perihelion distances, makes them vulnerable to planetary scattering. For trans-Neptunian objects with $a$ a little larger than $1600$~au (see Fig.~\ref{fig:zones}), the period of the orbital cycles induced by galactic tides counts in tens of Gyrs \citep{HIGUCHI-etal_2007}. This means that such objects do not even have the time to perform a complete cycle over the age of the solar system. However, in a $4.5$-Gyrs duration, the variation of perihelion distance can still reach several hundreds of astronomical units, as detailed in the next section.
   
\section{The borderline ``inert'' region}\label{sec:inert}
   In Sect.~\ref{sec:sec}, we studied the dynamics dominated by the secular perturbations from the planets. In Fig.~\ref{fig:zones}, this corresponds to the interval of $q$ located above the scattering region, for a semi-major axis small enough for the galactic tides to remain inefficient ($a\lesssim 500$~au), and between the mean-motion resonances. In Sect.~\ref{sec:gt}, we studied the dynamics dominated by the secular effects of the galactic tides. In Fig.~\ref{fig:zones}, this corresponds to the interval of $q$ located above the diffusive region, for a semi-major axis high enough for the planetary perturbations to remain inefficient ($a\gtrsim 1600$~au).
   
   In between ($500\lesssim a\lesssim 1600$~au), there necessarily exists an intermediate regime where perturbations from the planets and from the galactic tides have the same order of magnitude. It marks the dynamical frontier between the Kuiper-belt and the Oort-cloud populations. This region has been recently studied by \cite{SAILLENFEST-etal_2019}. They dubbed it the ``inert Oort cloud'' because both types of perturbations are small in this region, leading to orbits that are expected to be frozen in time (or ``fossilised''). Interestingly, \cite{SAILLENFEST-etal_2019} found that the truly inert region is actually very small (see Fig.~\ref{fig:zones}, where it is represented in blue). Between the two branches of the inert region, the combined action of the planets and of the galactic tides produces quite substantial variations of the orbital inclination and perihelion distance over the age of the solar system. In this section, we recall the main characteristics of this dynamics.
   
   As before in this review article, the computations below only include the known planets of the solar system. The effects of a hypothetical unobserved planet (and in particular, the ``Planet 9'' proposed by \citealp{BATYGIN-BROWN_2016} and studied by many authors afterwards) are not taken into account. The existence of Planet 9 would dramatically change the dynamical structure described in this section, and turn the weakly-perturbed intermediate regime between the Kuiper belt and the Oort cloud into a very active dynamical region. These aspects are further discussed in Sect.~\ref{sec:sculpt}.
   
\begin{figure}
      \centering
      \includegraphics[width=0.7\columnwidth]{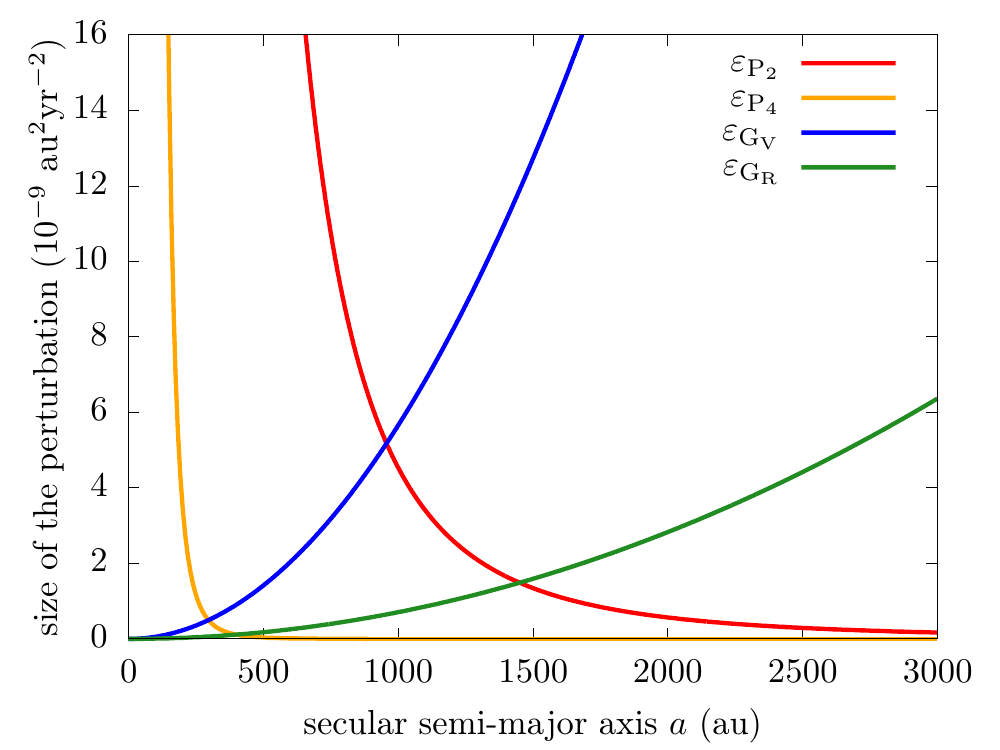}
      \caption{Size of the small parameters listed in Eqs.~\eqref{eq:epsP} and \eqref{eq:epsG} with respect to the secular semi-major axis of the small body. Adapted from \cite{SAILLENFEST-etal_2019}.}
      \label{fig:eps}
   \end{figure}

   For now, we consider the orbital evolution of a small body perturbed both by the known planets and by the galactic tides, in a region where mean-motion resonances with the planets have a negligible effect. The secular Hamiltonian function can be directly obtained from Sects.~\ref{sec:sec} and \ref{sec:gt}:
   \begin{equation}
      \mathcal{M}_\mathrm{PG} = \varepsilon_{\mathrm{P}_2}\mathcal{M}_{\mathrm{P}_2} + \varepsilon_{\mathrm{P}_4}\mathcal{M}_{\mathrm{P}_4} + \varepsilon_{\mathrm{G}_\mathrm{V}}\mathcal{M}_{\mathrm{G}_\mathrm{V}} + \varepsilon_{\mathrm{G}_\mathrm{R}}\mathcal{M}_{\mathrm{G}_\mathrm{R}} \,,
   \end{equation}
   where the expression of each part is given in Eqs.~\eqref{eq:Fp} and \eqref{eq:FGvr}. The explicit expressions of the small parameters (see Eqs.~\ref{eq:epsP} and \ref{eq:epsG}) have been chosen such that the Hamiltonian functions $\mathcal{M}_{\mathrm{P}_2}$, $\mathcal{M}_{\mathrm{P}_4}$, $\mathcal{M}_{\mathrm{G}_\mathrm{V}}$, and $\mathcal{M}_{\mathrm{G}_\mathrm{R}}$ have the same order of magnitude for $e=0$. The secular semi-major axis rules the relative importance of the different perturbation terms. Figure~\ref{fig:eps} shows that below $a\sim 600$~au, the planetary perturbations dominate over the galactic tides by more than a factor 10. The situation is reversed beyond $a\sim 1500$~au. In between, both kinds of perturbations have the same order of magnitude ($\varepsilon_{\mathrm{P}_2}$ and $\varepsilon_{\mathrm{G}_\mathrm{V}}$ cross at $a\sim 950$~au). However, since the eccentricity appears at the denominator in $\mathcal{M}_{\mathrm{P}_2}$ (see Eq.~\ref{eq:Fp}), the planetary perturbations always dominate in the high-eccentricity regime.
   
\begin{figure}
      \includegraphics[width=0.49\columnwidth]{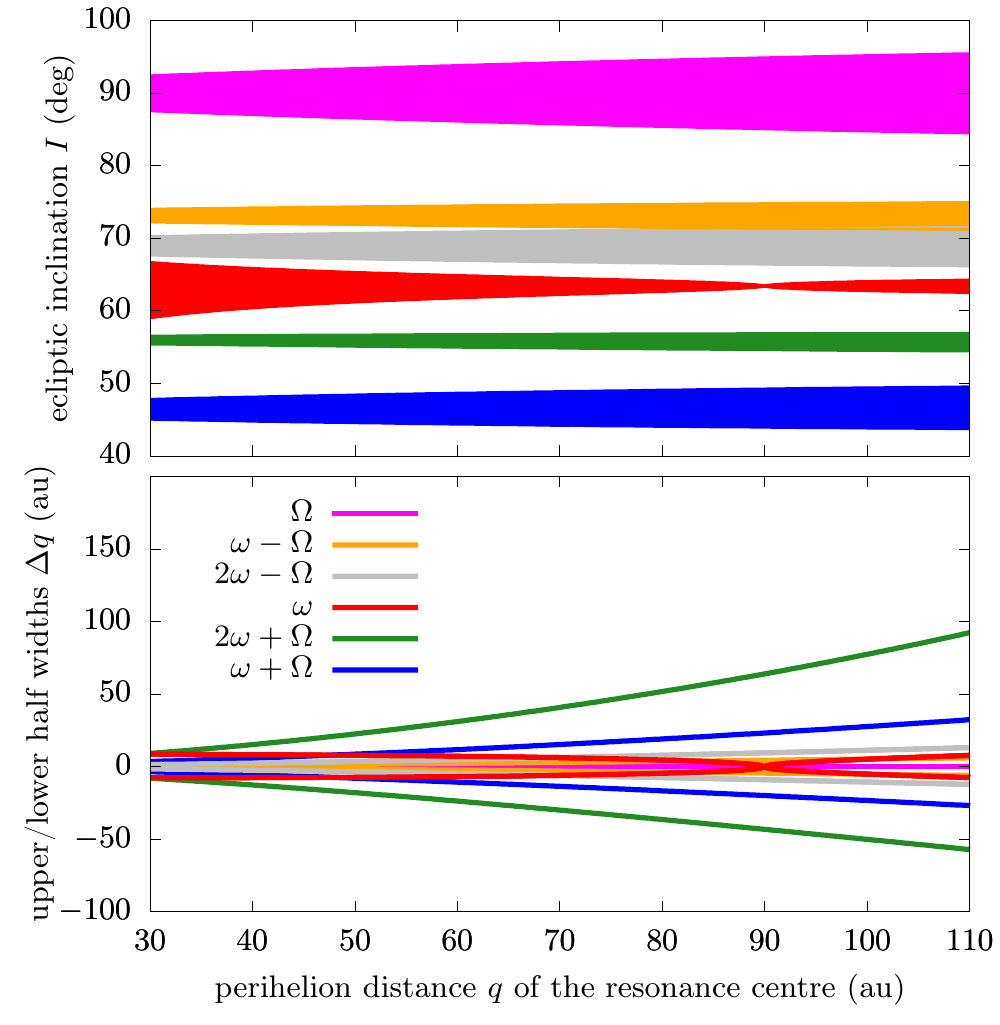}
      \hfill
      \includegraphics[width=0.49\columnwidth]{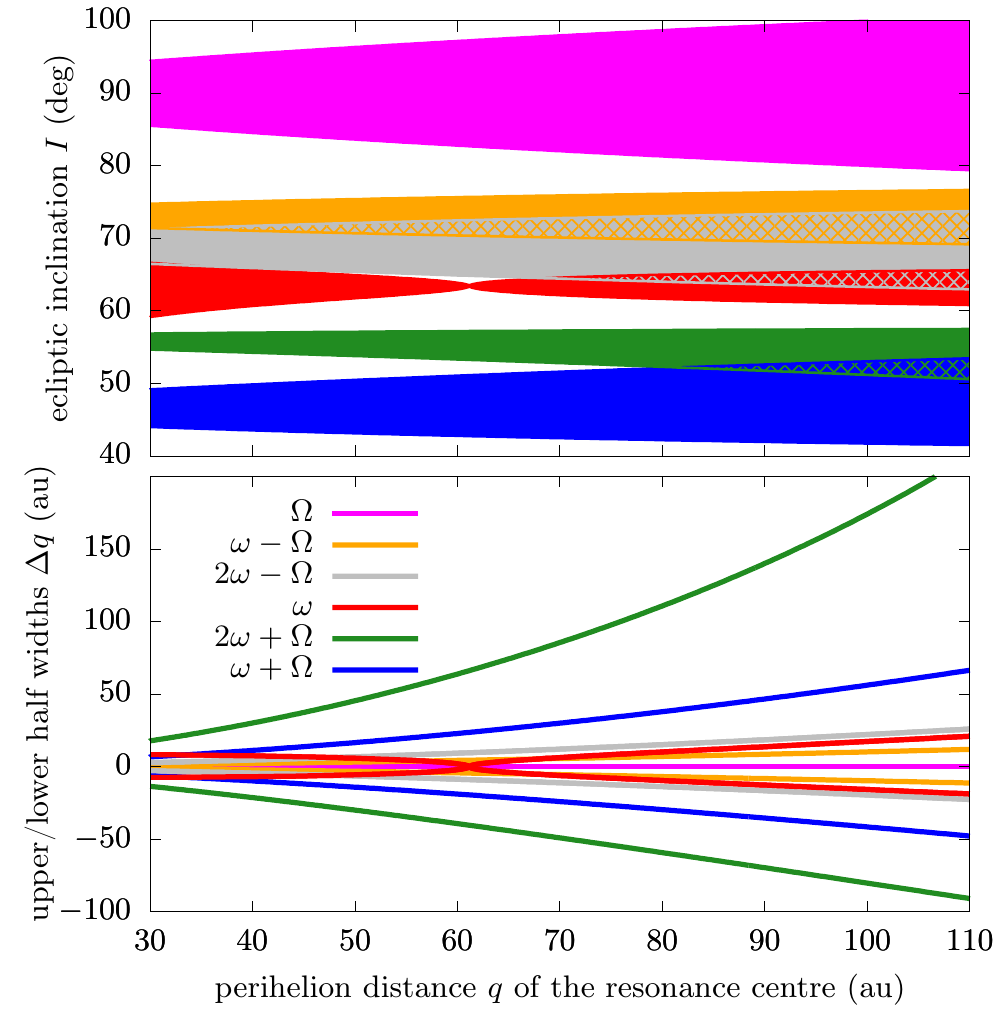}
      \caption{Location and widths of the strongest resonances in the planetary regime weakly perturbed by the galactic tides. The semi-major axis taken as parameter is $a=500$~au (left) and $a=700$~au (right). For better visibility, the perihelion distance of the resonance centre is directly used as horizontal axis. \emph{Top:} location and width in inclination (filled areas). \emph{Bottom:} upper and lower half widths in perihelion distance, for a centre given by the horizontal axis. The hatched regions mean overlap.}
      \label{fig:secres}
   \end{figure}

   As shown in Sects.~\ref{sec:sec} and \ref{sec:gt}, the secular dynamics is integrable in the small semi-major axis regime (where the planetary perturbations strongly dominate, i.e. $\varepsilon_\mathrm{P}\gg\varepsilon_\mathrm{G}$), as well as in the high semi-major axis regime (where the galactic tides strongly dominate, i.e. $\varepsilon_\mathrm{G}\gg\varepsilon_\mathrm{P}$). In order to study the dynamics in the intermediate regime, we must express the Hamiltonian function using a unique set of coordinates. We write $\psi$ the inclination of the ecliptic plane with respect to the galactic plane, and $\alpha$ its ascending node. The precession of the ecliptic pole can safely be neglected over the age of the solar system, such that $\psi$ and $\alpha$ are constant angles. The ascending node of the ecliptic can therefore be used as the origin of longitudes in the galactic frame, meaning that $\alpha\equiv 0$. Using this convention, the change of coordinates between the ecliptic reference frame (Sect.~\ref{sec:sec}) and the galactic reference frame (Sect.~\ref{sec:gt}) is a simple rotation of angle $\pm\psi$ around the first axis. In particular, we have
   \begin{equation}\label{eq:cosI}
      \cos I = \cos\psi\cos I_\mathrm{G} + \sin\psi\cos\Omega_\mathrm{G}\sin I_\mathrm{G} \,.
   \end{equation}
   The other conversion formulas can be found in \cite{SAILLENFEST-etal_2019}. Expressed in the ecliptic reference frame, the Hamiltonian function $\mathcal{M}_{\mathrm{G}_\mathrm{V}}$ becomes
   \begin{equation}\label{eq:FGE}
      \begin{aligned}
         \mathcal{M}_{\mathrm{G}_\mathrm{V}} = \frac{-1}{32}&\Bigg[ 2 \, (3e^2 + 2)(2C^2\cos^2I + S^2\sin^2I - 2) \\
         &  - 8  \, CS         (3e^2 + 2) \cos I\sin I       \, \cos(\Omega) \\
         &  + 5 \, S^2        e^2        (\cos I + 1)^2                \, \cos(2\omega + 2\Omega) \\
         &  + 20  \, CS         e^2        (\cos I + 1)\sin I \, \cos(2\omega + \Omega) \\
         &  + 10 \, (3C^2 - 1) e^2       \sin^2I                        \, \cos(2\omega) \\
         &  + 20  \, CS         e^2        (\cos I - 1)\sin I \, \cos(2\omega - \Omega) \\
         &  + 5 \, S^2        e^2        (\cos I - 1)^2                \, \cos(2\omega - 2\Omega) \\
         &  + 2 \, S^2        (3e^2 + 2) \sin^2I                       \, \cos(2\Omega)
         \Bigg] \,,
      \end{aligned}
   \end{equation}
   where $C\equiv\cos\psi$ and $S\equiv\sin\psi$. As in Sect.~\ref{sec:gt}, we will neglect the radial component of the galactic tides, dropping one degree of freedom. However, contrary to \cite{SAILLENFEST-etal_2019}, we will keep the hexadecapolar planetary contribution in order to resolve better the regime of low semi-major axes, where $\varepsilon_{\mathrm{P}_4} \gtrsim \varepsilon_{\mathrm{G}_\mathrm{V}}$ (see Fig.~\ref{fig:eps}). Hence we study the dynamics driven by the Hamiltonian:
   \begin{equation}\label{eq:Fpg}
      \mathcal{M} = \varepsilon_{\mathrm{P}_2}\mathcal{M}_{\mathrm{P}_2} + \varepsilon_{\mathrm{P}_4}\mathcal{M}_{\mathrm{P}_4} + \varepsilon_{\mathrm{G}_\mathrm{V}}\mathcal{M}_{\mathrm{G}_\mathrm{V}} \,.
   \end{equation}
   For small-enough semi-major axes and/or small-enough perihelion distances, the term $\varepsilon_{\mathrm{P}_2}\mathcal{M}_{\mathrm{P}_2}$ is strongly dominant in the overall Hamiltonian $\mathcal{M}$. This means that the remaining terms of the Hamiltonian can be treated using a perturbative approach. Such an approach is greatly eased by the fact that $\varepsilon_{\mathrm{P}_2}\mathcal{M}_{\mathrm{P}_2}$ is already expressed in action-angle coordinates (see Eq.~\ref{eq:Fp}). Hence, we can directly study the effect of each term of the perturbation $\varepsilon_{\mathrm{P}_4}\mathcal{M}_{\mathrm{P}_4} + \varepsilon_{\mathrm{G}_\mathrm{V}}\mathcal{M}_{\mathrm{G}_\mathrm{V}}$. 
   \begin{itemize}
      \item The first term of Eq.~\eqref{eq:FGE} does not not contain the angles; it acts therefore only as a small modulation of the precession velocities $\dot{\omega}$ and $\dot{\Omega}$ governed by $\varepsilon_{\mathrm{P}_2}\mathcal{M}_{\mathrm{P}_2}$, given at Eq.~\eqref{eq:omegadot}. The same holds for the first term of $\mathcal{M}_{\mathrm{P}_4}$ (see Eq.~\ref{eq:Fp}).
      \item The second term of Eq.~\eqref{eq:FGE} is factored by $\cos\Omega$. Strictly speaking, this term cannot be called ``resonant'' because it features no separatrix. As shown by \cite{SAILLENFEST-etal_2019}, this term is responsible for the emergence of a ``Laplace plane'' analogous to the one found in the satellite case: the orbit does not precess about the ecliptic pole, as it would for $\varepsilon_{\mathrm{G}_\mathrm{V}}=0$, but about a tilted pole. For very large values of the semi-major axis, this tilted pole tends to be the galactic pole.
      \item All the remaining terms of Eq.~\eqref{eq:FGE} correspond to resonances and libration zones for $\omega$ and $\Omega$. In particular, we note that there is a term factored by $\cos(2\omega)$, which adds to the one coming from the hexadecapolar term of the planetary perturbations $\varepsilon_{\mathrm{P}_4}\mathcal{M}_{\mathrm{P}_4}$ (see Eq.~\ref{eq:Fp}). The dynamics in the vicinity of each resonance can be studied using the pendulum approximation, giving analytical expressions of the resonance widths.
   \end{itemize}
   Figure~\ref{fig:secres} shows the location and widths of all the strongest resonances (the ones that directly appear in the Hamiltonian). This figure is restricted to small perihelion distances, for the planets to remain by far the dominant term of the dynamics. We focus on prograde orbits, since the resonances for $I>90^\text{o}$ are obtained by replacing $\cos I$ by $-\cos I$ and $\Omega$ by $-\Omega$. As shown in the top panels of Fig.~\ref{fig:secres}, the libration zone of $\Omega$ has by far the largest width in inclination. As shown in the bottom panels of Fig.~\ref{fig:secres}, the resonances $\omega+\Omega$ and $2\omega+\Omega$ are by far the largest ones in perihelion distance. The other resonances are quite small in comparison, and the libration zone of $\Omega$ even has a null width in~$q$. The libration zone of $\omega$ at $I\approx 63^\text{o}$ is due both to the hexadecapolar planetary term $\varepsilon_{\mathrm{P}_4}\mathcal{M}_{\mathrm{P}_4}$ and to the galactic term $\varepsilon_{\mathrm{G}_\mathrm{V}}\mathcal{M}_{\mathrm{G}_\mathrm{V}}$. Interestingly the two contributions have different signs, such that they cancel for some value of the perihelion distance. Beyond this value, the galactic contribution dominates, and we retrieve the graphs of \cite{SAILLENFEST-etal_2019}, for which $\varepsilon_{\mathrm{P}_4}\mathcal{M}_{\mathrm{P}_4}$ was neglected.
   
\begin{figure}
      \centering
      \includegraphics[width=\textwidth]{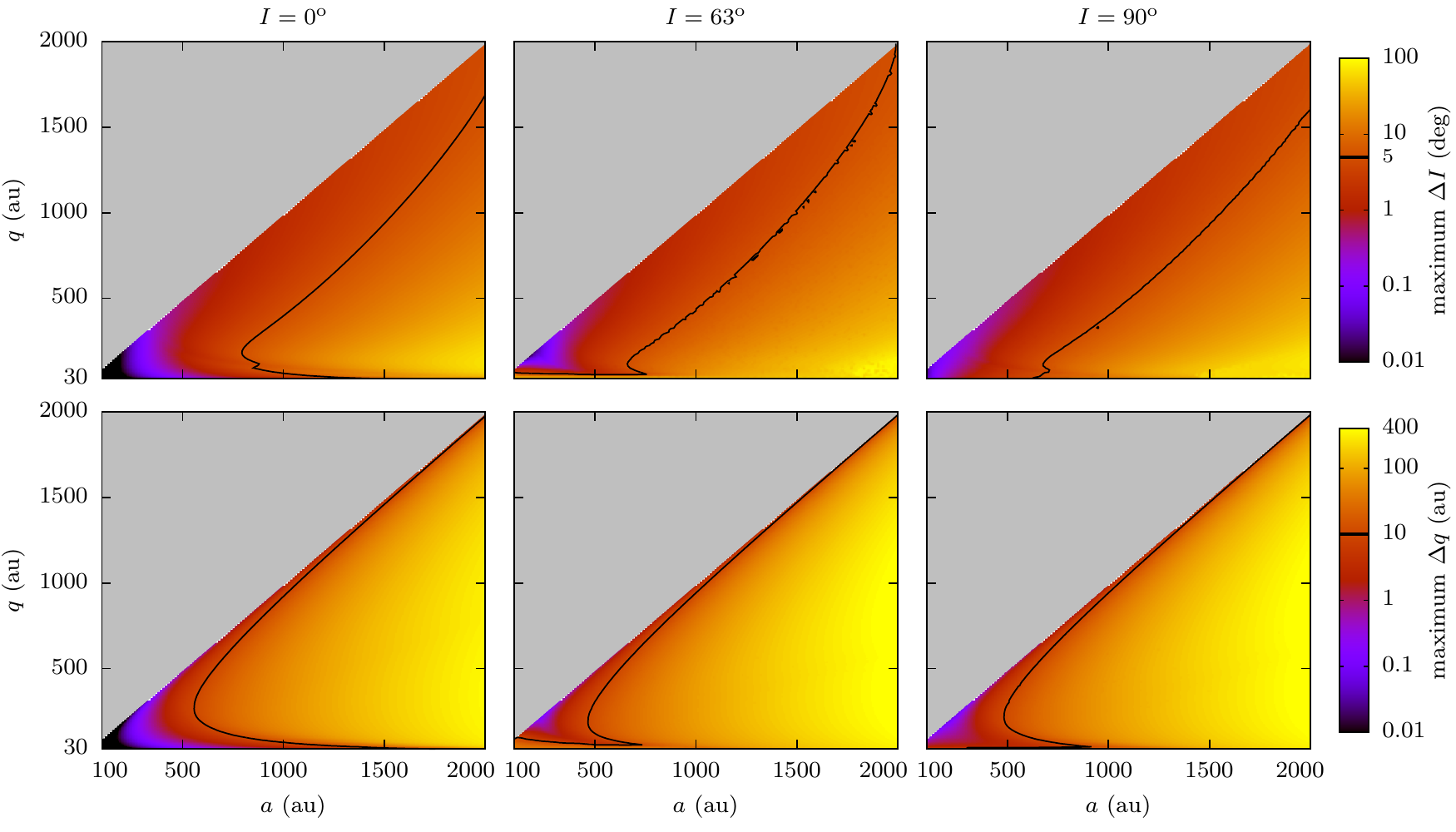}
      \caption{Limits of the inert region in the $(a,q,I)$ space. Each column corresponds to a different value of the inclination (see titles). The colour scale represents the maximum orbital variations reachable in $4.5$~Gyrs for initial conditions $(a,q,I)$. The top row shows the variations of ecliptic inclination, the bottom row shows the variations of perihelion distance (see labels on the right). The black level corresponds to a variation of $5^\text{o}$ in inclination (top row) or $10$~au in perihelion distance (bottom row). Below the black level, the region can be considered inert. Contrary to \cite{SAILLENFEST-etal_2019}, the hexadecapolar planetary term is included.}
      \label{fig:aqmap}
   \end{figure}

\begin{figure}
      \centering
      \includegraphics[width=\textwidth]{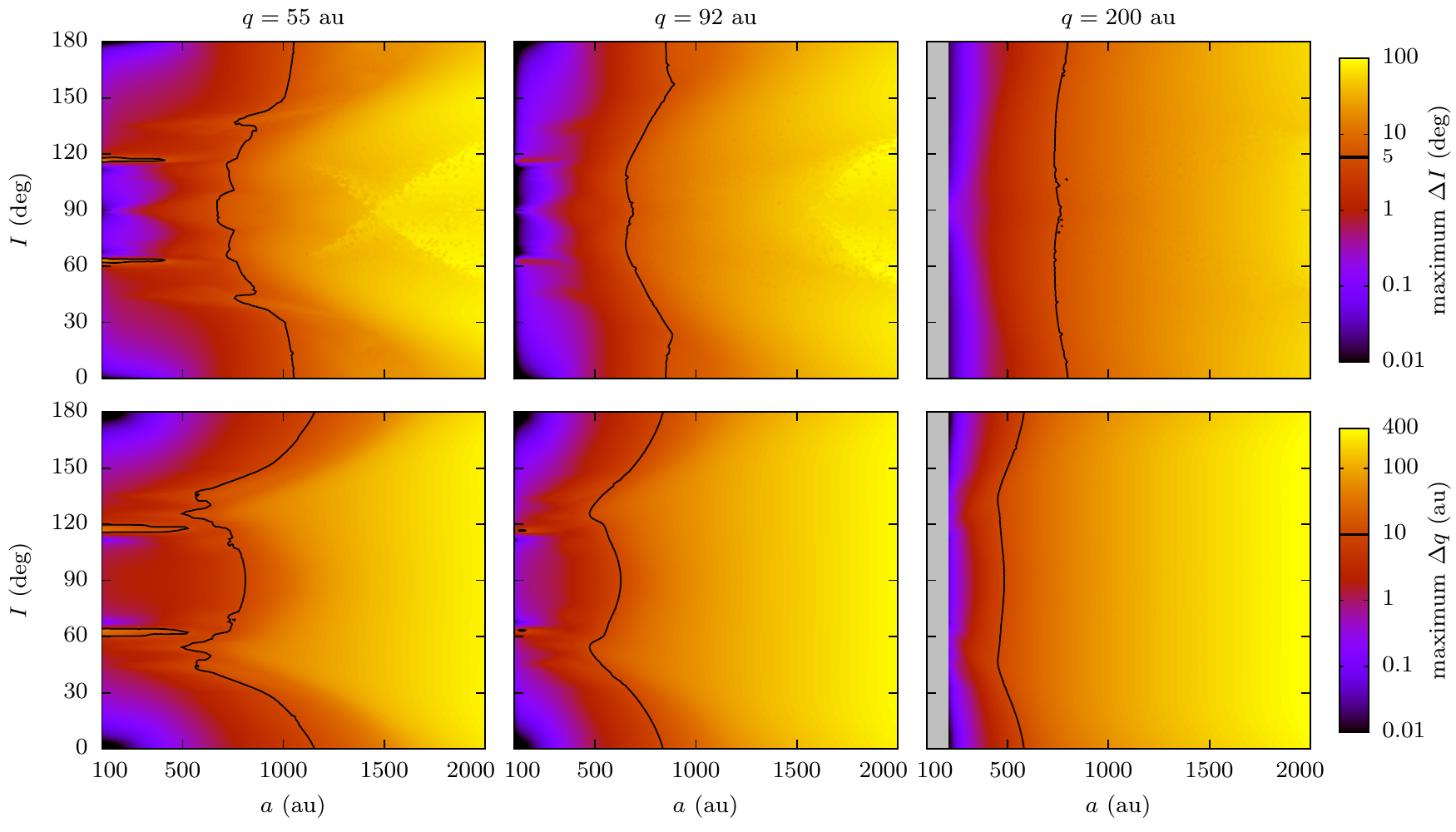}
      \caption{Same as Fig.~\ref{fig:aqmap} but in the $(a,I)$ plane. Each column corresponds to a different value of the perihelion distance (see titles).}
      \label{fig:aimap}
   \end{figure}

   As shown by Fig.~\ref{fig:secres}, when we increase the semi-major axis or the perihelion distance, the resonances become very large and overlap massively. For overly large resonances, the whole dynamical structure outlined with the perturbative approach is actually destroyed: the galactic tides cannot be treated anymore as a small perturbation of the quadrupolar term $\varepsilon_{\mathrm{P}_2}\mathcal{M}_{\mathrm{P}_2}$. \cite{SAILLENFEST-etal_2019} turned therefore to Poincar{\'e} sections. They showed that for $800\lesssim a\lesssim 1100$~au, a wide chaotic sea covers almost all the eccentricity range. Contrary to the scattering effect described in Sect.~\ref{sec:adiff}, this chaos is restricted to the secular system and only comes out on the long-term evolution. The diffusion timescales are large, but not to the point of being indiscernible in a $4.5$-Gyrs duration: the perihelion distance can actually vary from tens to hundreds of astronomical units, and the inclination can vary by tens of degrees. Figures~\ref{fig:aqmap} and \ref{fig:aimap} show the largest orbital changes reachable in $4.5$~Gyrs in the $(a,q,I)$ space using the Hamiltonian $\mathcal{M}$ from Eq.~\eqref{eq:Fpg}. The black curve delimits the ``inert''
   portion of the space, defined arbitrarily as $\Delta q<10$~au or $\Delta I<5^\text{o}$. In Fig.~\ref{fig:aqmap}, we retrieve the thin inert region illustrated in Fig.~\ref{fig:zones}. From Fig.~\ref{fig:aimap}, it is clear that the precise limit of the inert region depends on the inclination as well. We recognise the resonant structure displayed in Fig.~\ref{fig:secres}, and the overlapping of all resonances for growing $a$. These resonances ease the orbital variations of $q$ and $I$. For small semi-major axes and small perihelion distances, the effect of the hexadecapolar planetary term is clearly visible, in slightly enhancing the orbital variations at $I\approx 63^\text{o}$ and $117^\text{o}$. Apart from this feature, the limits of the inert region are the same as those computed by \cite{SAILLENFEST-etal_2019}. Two observed objects are located in the inert region: (90377) Sedna and 2012\,VP$_{113}$; and one observed object is located at its very border: 2015\,TG$_{387}$. Because of their highly eccentric orbits, these bodies cannot have formed in their current location. Their inert state implies that they remained ``fossilised'' there since the dramatic event that shaped their orbits. Discussions about their origin can be found in Sects.~\ref{sec:stars} and \ref{sec:sculpt}.
   
\section{Passing stars}\label{sec:stars}
   The solar system is surrounded by stars, whose individual gravitational attraction adds to the galactic tides described in Sect.~\ref{sec:gt}. The motion of stars is composed of a mean circling over the galactic centre, plus a peculiar velocity. In the neighbourhood of the sun, this peculiar velocity can be considered as isotropically distributed among stars \citep{RICKMAN-etal_2004,RICKMAN-etal_2005}. Even though the mean star-to-star distance is large compared to the extent of planetary systems, some stars happen to come close enough to stand out from the general galactic field, especially as seen from small bodies orbiting the sun on far-away trajectories. As recalled by \cite{COLLINS-SARI_2010}, the attraction of stars is already partly taken into account in the definition of the galactic tidal forces, and one must not count it twice. Therefore, the passing stars studied in this section refer to sporadic events, leading to impulsive changes in orbital elements that are qualitatively very different from the long-term effects discussed in Sect.~\ref{sec:gt}. \cite{RICKMAN-etal_2005} found that a maximum distance of about $10^6$~au for considering stars as individual objects is a reasonable limit. As we will show below, more than a million stars passed within this range over the age of the solar system.
   
   Looking at Fig.~\ref{fig:zones}, one can think of the effect of a passing star as ``shaking the box'', that is, producing a sudden spread of small bodies in the $(a,q,I)$ space. For a substantial shake, some fraction of small bodies can therefore be transferred abruptly in a distinct dynamical region. The impulses are weaker for Kuiper-belt objects than for more distant Oort cloud comets, but still quite noticeable \citep{SHEPPARD-etal_2019}. Different sequences of star passages only affect the timing and the efficiency of the spreading in the $a$, $q$, and $I$ directions, while the qualitative effect remains the same. Hence, contrary to the previous sections, the problems here are rather about the methods used than about the dynamics itself: \emph{i)} how to efficiently simulate the orbital perturbation due to a passing star, and \emph{ii)} how to build a realistic star sample. These questions are addressed in Sects.~\ref{sec:CIA} and \ref{sec:sample}. In Sect.~\ref{sec:stareffect}, we go a little more into details about the effects of stars for Kuiper-belt objects.
   
   \subsection{Perturbation by a single stellar passage}\label{sec:CIA} 
   In a heliocentric reference frame, the acceleration of a small body located in position $\mathbf{r}=(x,y,z)^\mathrm{T}$ due to the presence of a star with gravitational parameter $\mu_\star$ and position $\mathbf{r}_\star$ is:
   \begin{equation}\label{eq:star}
      \mathbf{F}_\star = -\mu_\star\left(\frac{\mathbf{r}-\mathbf{r}_\star}{\|\mathbf{r}-\mathbf{r}_\star\|^3} + \frac{\mathbf{r}_\star}{\|\mathbf{r}_\star\|^3}\right) \,.
   \end{equation}
   The first term is the direct acceleration, and the second one comes from the acceleration of the sun due to attraction of the passing star. In order to compute the total effect of the star on the orbit of the small body, we must integrate this acceleration over time, from the moment the star emerges from the mean galactic field, until the moment it reintegrates it. \cite{RICKMAN-etal_2005} found that a good compromise is to consider the path of the star located at $\|\mathbf{r}_\star\|<10^6$~au. The most accurate results are obtained by integrating numerically the equations of motion using sophisticated N-body codes and adding Eq.~\eqref{eq:star} to the acceleration of the small body. This method is best when studying the detailed outcome of some specific stellar passage, but it is particularly inefficient when dealing with thousands of star passages accumulated over the age of the solar system on a sample of millions of small bodies. This is why the classical impulse approximation and its variants have been developed and are still widely used today. These approximations are also useful to draw a qualitative picture of the outcome of stellar encounters, as detailed below.
   
   As mentioned by \cite{RICKMAN-etal_2005}, the classical impulse approximation was already used by \cite{OPIK_1932} in very early works about stellar passages, but it only became popular when the concept of Oort cloud was introduced and that the attention grew about its dynamics (see Sect.~\ref{sec:hist}). Following its reformulation by \cite{RICKMAN_1976}, numerous authors still stick to classical form of the impulse approximation, due to its simplicity and its analyticity (see e.g. \citealp{KAIB-etal_2011}, \citealp{HIGUCHI-KOKUBO_2015}, \citealp{TORRES-etal_2019}, and the references given by \citealp{RICKMAN-etal_2005}). We recall here its main characteristics and discuss its advantages over other more sophisticated methods.
   
   The classical impulse approximation relies on two assumptions made over the duration of the stellar encounter:
   \begin{enumerate}
      \item The star moves in a straight line with constant velocity.
      \item The sun and the small body are at rest.
   \end{enumerate}
   These assumptions especially hold for distant small bodies, because they have small orbital velocities, and they are mainly sensible to stars when they are at aphelion, where they move the slowest. The typical heliocentric velocity of a star is $50$~km$\cdot$s$^{-1}$ (see \citealp{RICKMAN-etal_2008}, who noted that the value cited by \citealp{RICKMAN-etal_2004,RICKMAN-etal_2005} is too small), while a small body with $a = 800$~au and $q = 40$~au has a velocity of about $0.2$~km$\cdot$s$^{-1}$ at aphelion. The velocity ratio is not as large as for Oort cloud comets, but it is still large enough for the classical impulse method to give satisfactory results in a statistical sense, unless in dramatic cases of very slow (but very improbable) encounters. Using these two assumptions, Eq.~\eqref{eq:star} can be integrated analytically from time $-\infty$ to $+\infty$, giving the total change of the velocity vector of the small body due to the stellar passage. Splitting the vector $\mathbf{r}_\star$ into the components that are parallel and perpendicular to the star track, we obtain
   \begin{equation}\label{eq:CIA}
      \Delta\mathbf{v} = \frac{2\mu_\star}{v_\star}\left(\frac{\mathbf{b}}{b^2} - \frac{\mathbf{b}_\odot}{b_\odot^2}\right) \,,
   \end{equation}
   where $v_\star$ is the constant velocity of the star, $\mathbf{b}$ is the vector pointing from the small body towards the closest position of the star along its track, $b\equiv \|\mathbf{b}\|$, and the $\odot$ index refers to the same quantities for the sun. Adopting a reference frame for which the $x$-axis is anti-parallel to the velocity vector of the star and the $y$-axis is parallel to $\mathbf{b}_\odot$, the three components of the velocity impulse are
   \begin{equation}\label{eq:CIAcoo}
      \left\{
      \begin{aligned}
         \Delta v_x &= 0 \,,\\
         \Delta v_y &= \frac{2\mu_\star}{v_\star}\left(\frac{b_\odot-y}{b^2} - \frac{1}{b_\odot}\right) \,,\\
         \Delta v_z &= -\frac{2\mu_\star}{v_\star}\frac{z}{b^2} \,.
      \end{aligned}
      \right.
   \end{equation}
   In these coordinates, we also have the relation
   \begin{equation}\label{eq:rel}
      b^2 = (y-b_\odot)^2 + z^2 \,.
   \end{equation}
   The velocity impulse can then be converted into the corresponding changes of orbital elements, as detailed for instance by \cite{RICKMAN_1976} or \cite{HIGUCHI-KOKUBO_2015}. Equation~\eqref{eq:CIA} clearly shows that the orbital change of the small body is due to a tidal effect, that is, to the difference of the attraction felt by the small body and the attraction felt by the sun. From Eq.~\eqref{eq:CIAcoo}, we also note that the impulse is perpendicular to the star track, and that it is larger if the star is slow and passes close.
   
   As remarked by \cite{RICKMAN-etal_2005}, however, very close passages invalidate assumption 1, and very slow passages invalidate assumption 2. This prompted \cite{DYBCZYNSKI_1994} to develop an improved variant of the impulse approximation in which the star follows an arbitrary hyperbolic orbit around the Sun. Considering again an infinite time before and after the encounter, the improved velocity change is
   \begin{equation}\label{eq:DIAcoo}
      \left\{
      \begin{aligned}
         \Delta v_x &= -\frac{2\mu_\star}{v_\star}\left(\frac{a}{c^2} - \frac{a_\odot}{c_\odot^2}\right) \,,\\
         \Delta v_y &= \frac{2\mu_\star}{v_\star}\left(\frac{b_\odot-y}{c^2} - \frac{b_\odot}{c_\odot^2}\right) \,,\\
         \Delta v_z &= -\frac{2\mu_\star}{v_\star}\frac{z}{c^2} \,,
      \end{aligned}
      \right.
   \end{equation}
   where $c^2 = a^2+b^2$ (and the same with $\odot$ index). This time, $v_\star$ represents the velocity of the star at infinity, and $a_\odot=(\mu+\mu_\star)/v_\star^2$ is the semi-major axis of the hyperbolic orbit of the star around the sun. Likewise, $a=\mu_\star/v_\star^2$, and the relation given at Eq.~\eqref{eq:rel} still holds. We retrieve the classical impulse approximation from Eq.~\eqref{eq:CIAcoo} by putting $a=a_\odot=0$. Dybczy{\'n}ski's impulse formula is still remarkably simple, and it allows us to completely drop assumption~1.
   
   Independently, \cite{EGGERS-WOOLFSON_1996} developed a sequential method in which the classical impulse from Eq.~\eqref{eq:CIA} is divided into several sub-impulses. This allowed them to account for the heliocentric motion of the small body during the stellar passage, that is, to completely drop assumption~2. These sub-impulses resemble the variable steps of a numerical integrator, but since they are directly set in terms of the true anomaly, the algorithm is still orders of magnitude faster than (non-regularised) numerical integrations.
   
   Eventually, both variants were put together by \cite{RICKMAN-etal_2005}, who dropped both assumptions. It was also shown that the approximation of neglecting the action of the galactic tides on the stellar trajectory in the timespan of its encounter with the sun is perfectly viable, and does not need any improvement \citep{LETO-etal_2007}. We will not enter into more details here, since these are only a matter of implementation technicalities. \cite{RICKMAN-etal_2005} found that the errors produced by the classical and modified impulse approximations are symmetrically distributed around zero. This means that, while they do not give the accurate outcome of an individual star passage, they still yield very satisfactory results in a statistical sense, when considering many star passages acting on many objects. Moreover, the impulse approximations are worst when the orbital variations induced are small, that is, when they matter less. For large orbital changes, even the errors using the classical impulse from Eq.~\eqref{eq:CIAcoo} rarely exceed 10\% for Oort cloud bodies ($a\gtrsim 10^4$~au). The situation gets worst for smaller semi-major axes like the ones shown in Fig.~\ref{fig:zones}, and \cite{FOUCHARD-etal_2007b} advocate using the sequential method whenever precise quantitative results are needed. Nowadays, studies requiring accurate statistics mostly use the sequential variant of \cite{RICKMAN-etal_2005} for large samples of small bodies. For instance, we can mention the simulations by \cite{FOUCHARD-etal_2017} that each include $10^7$ bodies. Thanks to the increasing power of computers, direct numerical integrations are sometimes preferred, but still at the cost of the sample size (for instance, \citealp{NESVORNY-etal_2017}, and \citealp{VOKROUHLICKY-etal_2019}, simulated $10^6$ bodies ``only'').
   
   \subsection{Building the stellar sample}\label{sec:sample}
   Having chosen a suitable method for modelling the effects of stars, one must build a sequence of star passages that will affect the orbit of small bodies over the timespan needed. Due to the fast motion of nearby stars, numerical integrations can be used to predict their trajectories only in a restricted timespan, of the order of $\pm 10$~Myr centred at present \citep{GARCIASANCHEZ_2001,TORRES-etal_2019}. Using this method, the \emph{Gaia} catalogue is accurate enough to spot the neighbouring stars that produced or will produce substantial injections of comets into the observable region \citep{FOUCHARD-etal_2011b,BERSKI-DYBCZYNSKI_2016}. For realistic simulations featuring several stars, however, the sequence of passages obtained must be corrected for the incompleteness of the catalogue used. In any case, one must turn to statistical methods for longer durations. This can be realised by: \emph{i)} measuring the density of stars in the neighbourhood of the sun, with their masses and velocity dispersions, \emph{ii)} deducing the distribution of star passages in the vicinity of the sun, \emph{iii)} extrapolating these quantities over the whole history of the solar system. For accurate estimates, point \emph{ii} requires the numerical propagation of nearby stars within the galactic potential \citep{BAILERJONES_2015}.
   
   Most authors still use the statistics of star passages computed by \cite{GARCIASANCHEZ_2001}, but slightly updated quantities can be found in \cite{TORRES-etal_2019}. In a near future, however, the full \emph{Gaia} catalogue will be available for building complete refined statistics of stellar passages. New estimates of the total stellar encounter rate, corrected for incompleteness, can already be found in \cite{BAILERJONES_2018} and \cite{BAILERJONES-etal_2018}. The velocity vector $\mathbf{V}_\star$ of each star is measured with respect to its ``local standard of rest'' (LSR), that is, the reference frame that follows the average motion of galactic material in its neighbourhood. $\mathbf{V}_\star$ is called the ``peculiar velocity'' of the star, and it is directed towards the ``apex'' of the star. The star's heliocentric velocity is then $v_\star = \|\mathbf{V}_\star-\mathbf{V}_\odot\|$, where $\mathbf{V}_\odot$ is the peculiar velocity of the sun with respect to the star's LSR. Combining these data for many stars, \cite{GARCIASANCHEZ_2001} computed the velocity dispersion of 13 categories of stars, providing a catalogue as complete as possible. They estimated the encounter frequencies $f$ of stars both using results from the literature and their numerical integration over $\pm 10$~Myrs of the nearby \emph{Hipparcos} stars, corrected from incompleteness. The encounter frequency is defined such that the number of stars encountered during a timespan $\Delta t$ within a radius $D$ is
   \begin{equation}
      N = fD^2\Delta t \,.
   \end{equation}
   Assuming an isotropic distribution of peculiar velocities, \cite{RICKMAN-etal_2008} used these data to compute the mean heliocentric velocity of the encounters and its standard deviation for each category of stars. These quantities are gathered in Table~\ref{tab:stars}. One finds a total encounter frequency of $10.525$ star passages per Myr in a sphere of one parsec, corresponding to about $1800$ passages within $40\,000$~au in $4.5$~Gyrs, and about one passage within $1000$~au. Preliminary results from the \emph{Gaia} catalogue show about twice as many encounters \citep{BAILERJONES_2018,BAILERJONES-etal_2018}, but individual statistics by star category are still missing. As noted by \cite{GARCIASANCHEZ_2001}, the most frequent encounters are with low-mass and high-velocity stars. The massive and slow stars that are expected to perturb most the orbits of small bodies only represent a small fraction of all passages (even though their effects can be decisive in regulating the flux of long-period comets, see \citealp{FOUCHARD-etal_2011}).
   
   \begin{table}
      \begin{equation*}
         \begin{array}{l|l|l|r|r}
            \hline
            \text{Type} & <m_\star>\ (M_\odot) & f \text{ (pc$^{-2}\cdot$Myr$^{-1}$)} & <v_\star> \text{ (km$\cdot$s$^{-1}$)} & \sigma_\star \text{ (km$\cdot$s$^{-1}$)} \\
            \hline
            \hline
            \text{B0} & 9    & 0.005 & 24.6 &  6.7 \\
            \text{A0} & 3.2  & 0.03  & 27.5 &  9.3 \\
            \text{A5} & 2.1  & 0.04  & 29.3 & 10.4 \\
            \text{F0} & 1.7  & 0.15  & 36.5 & 12.6 \\
            \text{F5} & 1.3  & 0.08  & 43.6 & 15.6 \\
            \text{G0} & 1.1  & 0.22  & 49.8 & 17.1 \\
            \text{G5} & 0.93 & 0.35  & 49.6 & 17.9 \\
            \text{K0} & 0.78 & 0.34  & 42.6 & 15.0 \\
            \text{K5} & 0.69 & 0.85  & 54.3 & 19.2 \\
            \text{M0} & 0.47 & 1.29  & 50.0 & 18.0 \\
            \text{M5} & 0.21 & 6.39  & 51.8 & 18.3 \\
            \text{wd} & 0.9  & 0.72  & 80.2 & 28.2 \\
            \text{gi} & 4    & 0.06  & 49.7 & 17.5 \\
            \hline
         \end{array}
      \end{equation*}
      \caption{Stellar parameters as computed by \cite{GARCIASANCHEZ_2001} and \cite{RICKMAN-etal_2008}. The first column gives the type of the star, with ``wd'' for white dwarfs and ``gi'' for giants; the second column gives the average mass of the star according to its type; the third column gives the encounter frequency in number per Myr within a sphere of $1$~parsec; the fourth and fifth columns give the mean heliocentric encounter velocity and its standard deviation.}
      \label{tab:stars}
   \end{table}
   
   Using the values from Table~\ref{tab:stars}, one can compute a sample of star passages that is statistically similar to the ones encountered by the sun in a given interval of time. We refer to \cite{RICKMAN-etal_2008} for the complete procedure. A slight improvement was added by \cite{VOKROUHLICKY-etal_2019}, who considered also a statistical distribution of the masses for each star category instead of simply picking the average mass given in Table~\ref{tab:stars} (this process was restricted to B0 stars in \citealp{RICKMAN-etal_2008}). For simplicity, most authors consider that the current parameters given in Table~\ref{tab:stars} are constant over the age of the solar system. However, a few studies have been dedicated to the variations of these quantities as the sun migrates vertically and radially (see e.g. \citealp{KAIB-etal_2011}). In particular, \cite{MARTINEZBARBOZA-etal_2017} found that according to the precise path of the sun through the Galaxy, the total encounter frequency in $4.6$~Gyrs can differ by $\pm 50\%$ with respect to the central value given by Table~\ref{tab:stars}. 
   
   \subsection{Effects of stars on Kuiper-belt objects}\label{sec:stareffect}
   Using the tools detailed above, \cite{RICKMAN-etal_2004} studied the cumulative effect of passing stars on the orbital distribution of trans-Neptunian objects with $a\lesssim 1000$~au, that is, the region detailed in Fig.~\ref{fig:zones}. They found that over $4$~Gyrs, small bodies with $a\approx 500$~au and $q\approx 35$~au have a $40\%$ chance of receiving only small impulses producing negligible orbital variations, and a $60\%$ chance of receiving significant impulses producing a wide distribution of perihelion distance, extending almost up to $1000$~au (but with a sharp decrease in probability). As remarked by \cite{EGGERS-WOOLFSON_1996}, positive or negative increments in perihelion distance due to passing stars roughly have the same probability, but the absolute limit at $q=0$ leads to an asymmetric extended tail in the positive direction. This means that small bodies located in the bottom part of Fig.~\ref{fig:zones} and affected by planetary scattering are sporadically injected into the top part of the figure, and in particular into the inert zone where they are safely stored for billions of years. As confirmed by \cite{SHEPPARD-etal_2019}, however, each set of stellar encounters is unique and their cumulative effect is strongly dependent on the few most powerful passages, which produce most of the dispersion.
   
   \cite{RICKMAN-etal_2004} concluded that a stellar passage with a minimum heliocentric distance of $800$~au would naturally create inert objects like (90377) Sedna. This strengthened the result by \cite{MORBIDELLI-LEVISON_2004} that a star passage was the most likely scenario able to explain this kind of orbit. Even if such close encounters are expected to be very few since the sun left its birth cluster (less than one over $4.5$~Gyrs, see Sect.~\ref{sec:sample}), they cannot be totally ruled out statistically; however, close and slow encounters are much more likely to have happened when the sun was still part of its birth cluster (when $f$ was higher and $<v_\star>$ lower than quoted in Table~\ref{tab:stars}). Moreover, a late stellar passage at less than a few thousands astronomical units would have emptied the Oort cloud in a dramatic comet shower. This also favours a very early event, when the Oort cloud was not yet formed (see Sect.~\ref{sec:sculpt} and the review by \citealp{MORBIDELLI-NESVORNY_2019}).
   
   Additionally to stars, a few giant molecular clouds are expected to have passed by since the formation of the sun \citep{KOKAIA-DAVIES_2019}. A close encounter with a giant molecular cloud can produce a large variety of outcomes for different impact parameters and encounter velocities. But again, no very strong encounter could have happened after the formation of the Oort cloud, otherwise it would have been completely depleted.
   
\section{Conclusions: sculpting the trans-Neptunian populations}\label{sec:sculpt}
   In the previous sections, we have reviewed the dynamical mechanisms that are known to affect the orbits of trans-Neptunian objects. They involve numerous distinct classes of dynamics, such as short-term chaotic diffusion (Sect.~\ref{sec:adiff}), quasi-integrable non-resonant trajectories (Sects.~\ref{sec:sec} and \ref{sec:gt}), isolated resonances (Sects.~\ref{sec:res} and \ref{sec:inert}), long-term chaotic diffusion (Sect.~\ref{sec:inert}), and even statistical events (Sect.~\ref{sec:stars}).
   
   Below a threshold of perihelion distance ($q\lesssim 45$~au), the planetary scattering produces a chaotic diffusion of semi-major axis (Sect.~\ref{sec:adiff}). Beyond this threshold and for small semi-major axes ($q\gtrsim 45$~au, $a\lesssim 500$~au), the dynamics is governed by secular planetary perturbations (Sect.~\ref{sec:sec}), and isolated mean-motion resonances with the planets that are able to produce large-amplitude variations of the perihelion distance (Sect.~\ref{sec:res}). Beyond the scattering threshold and for moderate semi-major axes ($q\gtrsim 45$~au and $500\lesssim a\lesssim 1600$~au), the galactic tides combine with secular planetary perturbations, producing a wide chaotic zone for $q$ and $I$. However, the diffusion timescales are very long, and a large portion of this region can be considered inert  (Sect.~\ref{sec:inert}). For large semi-major axes ($a\gtrsim 1600$~au), the galactic tides dominate over secular planetary perturbations and produce large-amplitude eccentricity and inclination cycles, possibly carrying small bodies in and out of the scattering region (Sect.~\ref{sec:gt}). Finally, passing stars produce sporadic jumps of small bodies in the $(a,q,I)$ space, but they are efficient in the Kuiper belt only for close passages, that probably did not happen since little after the sun left its stellar birth cluster (Sect.~\ref{sec:stars}). The different dynamical regions are summarised in Fig.~\ref{fig:zones}, where their limits correspond to the current state of the solar system and its galactic environment. All these dynamical mechanisms are at play since the early stages of the planetary formation, after the dispersal of the circumsolar gas disc, some $4.5$~Gyrs ago. The question of how all of them contributed to sculpt the orbital distribution of small bodies, and what was the initial state of the solar system that led to the observed distributions, is a very active field of research. From the last two decades or so, a unified picture has started to emerge, linking all populations of small bodies through a single scenario. A summary of the last advances can be found in the recent review by \cite{MORBIDELLI-NESVORNY_2019}. We list below the key elements of this scenario in the context of the dynamical mechanisms described throughout this review paper.
   
   After the dispersal of the circumsolar gas disc, the giant planets were initially located much closer to the sun than today, and the proto-Kuiper belt extended from Neptune's orbit of that time (say, about $20$~au) to $50$~au, with a massive inner component ranging up to the current location of Neptune (about $30$~au). As the gas dissipated, planets cleared the vicinity of their orbits by scattering planetesimals away, which, by conservation of angular momentum, made them migrate radially \citep{FERNANDEZ-IP_1984,MALHOTRA_1993,MALHOTRA_1995,LEVISON-etal_2007}. A phase of instability was then triggered when two giant planets crossed a mean-motion resonance, leading to the ejection of an enormous quantity of planetesimals (and even possibly one of the giant planets themselves, see \citealp{NESVORNY-MORBIDELLI_2012}). This was probably the moment where most of the Oort cloud population was created\footnote{An early formation of the Oort cloud, during the planetary formation, is unlikely because gas drag prevents objects from being ejected onto such distant orbits \citep{BRASSER-etal_2007}.}, as galactic tides and passing stars lifted the perihelion of recently scattered bodies out the reach of planetary perturbations (see Sects.~\ref{sec:gt} and \ref{sec:stars}). Finally, by ejection of planetesimals that survived the instability, the giant planets roughly circularised again and ended their migration at their present-day locations. Hence, in this scenario, Neptune migrated outwards across the inner, massive portion of the proto-Kuiper belt, and even brutally outwards during the instability phase. This migration, combined with all the dynamical mechanisms described above, is thought to have sculpted the Kuiper belt into the different populations of trans-Neptunian objects that are observed today. The observed trans-Neptunian objects are generally divided into several distinct populations according to their current orbital state \citep{GLADMAN-etal_2008}:
   \begin{itemize}
      \item[$\bullet$] The ``classical Kuiper belt'' gathers objects that have roughly circular orbits located mainly between the $2:3$ and $1:2$ mean-motion resonances with Neptune ($42\lesssim a\lesssim 48$~au), but are not locked in resonance\footnote{Objects of the classical Kuiper belt are sometimes called ``Cubewanos'' in reference to their first observed member, 1992\,QB$_1$, now officially named (15760) Albion. See Sect.~\ref{sec:hist} for a historical perspective.}. This places them inside (though at the border) of the inert zone, meaning that their orbits remain virtually unchanged through time, apart from precession due to the secular effect of the giant planets (see Sect.~\ref{sec:sec}). The classical Kuiper belt is generally divided into two sub-categories: the ``cold'' and ``hot'' populations. The cold classical objects have orbital inclinations $I\lesssim 5^\text{o}$. They are thought to have been formed in situ, and have only been slightly affected by the migration of Neptune. This left them on orbits that are close to the disc-like structure expected from formation models \citep{EDGEWORTH_1949,KUIPER_1951}. The hot classical objects have somewhat larger orbital inclinations ($5^\text{o}\lesssim I\lesssim 30^\text{o}$), but their distribution partially overlap with the cold population. Their spectral properties actually differ from the cold population, indicating a different region of formation within the protoplanetary disc. They are thought to have been formed below $30$~au and have been scattered away by Neptune during its outward migration \citep{MORBIDELLI-NESVORNY_2019}. Before ending on their final stable orbits, they are hence expected to have undergone a complex combination of scattering and temporary captures in resonant lifts (see Sect.~\ref{sec:adiff}). Due to the ongoing migration of Neptune that shifted the resonance locations, they have eventually been released out of resonance in the inert zone (see Sect.~\ref{sec:res}).
      \item[$\bullet$] The ``resonant objects'' are the small bodies currently locked in mean-motion resonance with Neptune. They probably mainly come from the same source as the hot population (i.e. from below $30$~au), but contrary to hot classical objects they managed to adiabatically follow the resonances during the migration of Neptune, or they have been captured in resonance after the end of migration, such that the resonant link was not broken. As mentioned in Sect.~\ref{sec:res}, the number of objects that are currently locked in the different mean-motion resonances gives hints about the properties of Neptune's migration: a smooth and slow migration leads to crowded resonances (since all resonant bodies are steadily carried away within the resonances), whereas a grainy and fast migration leads to empty resonances (since the resonances are gone before even affecting bodies). A good compromise for explaining the observations seems to be a slow and grainy migration, as found by \cite{LAWLER-etal_2019}.
      \item[$\bullet$] Objects of the ``scattered disc'' have the same origin as the hot and resonant populations: they also formed below $30$~au and were scattered away by Neptune, but with the difference of never founding a stable parking orbit until today. Hence, they still wander about in the scattering region of Fig.~\ref{fig:zones}, following the dynamics described in Sect.~\ref{sec:adiff}. Their dynamics includes captures in mean-motion resonance with Neptune, that are mostly temporary, but that can also turn virtually permanent if the high-perihelion trapping mechanism is triggered (see Sect.~\ref{sec:res}). Their dynamics also includes injection paths to the inner regions of the solar system (production of centaurs and Jupiter-family comets, as it was understood long ago by \citealp{FERNANDEZ_1980}, \citealp{TORBETT_1989}, \citealp{DUNCAN-LEVISON_1997}), or ejection paths towards the Oort cloud (see e.g. \citealp{GABRYSZEWSKI-RICKMAN_2010}). In the latter case, objects can either be purely ejected from the solar system if the scattering is brutal enough, or the galactic tides can detach their orbit from the chaotic region before the ejection, making them members of the Oort cloud. However, since the trajectories driven by the galactic tides are quasi-periodic (see Sect.~\ref{sec:gt}), they will inevitably cycle back towards the scattered disc, unless their orbits are reshaped by a timely passing star (see Sect.~\ref{sec:stars}). When objects come back to the scattering region, they can become scattered-disc objects or centaurs with high inclinations \citep{KAIB-etal_2019}, that may further evolve into Halley-type comets \citep{LEVISON-etal_2006}.
      \item[$\bullet$] The ``detached'', or ``fossilised'' objects are located deep inside the inert region (see Sect.~\ref{sec:inert}). They are not affected by scattering nor isolated resonances, and are not distant enough for the galactic tides to substantially affect them. As such, they have extremely stable orbits. They were probably initially part of the scattered disc, but they are now totally disconnected from it. The most notable detached bodies (Sedna, 2012\,VP$_{113}$, and 2015\,TG$_{387}$), sometimes called ``Sednoids'', are completely out of the range of action of any mean-motion resonance with Neptune. This implies that the perihelion-lifting mechanism described in Sects.~\ref{sec:adiff} and \ref{sec:res}, even if coupled with Neptune's migration, cannot explain their orbits. Hence, we must invoke external perturbers, and a close star passage is the most promising scenario (see Sect.~\ref{sec:stars}). A close star passage can either detach their orbits from the scattered disc, or deliver them into the solar system from the star's own planetary system \citep{RICKMAN-etal_2004,MORBIDELLI-LEVISON_2004,KENYON-BROMLEY_2004,JILKOVA-etal_2015}. However, Sedna-type orbits can only be produced through very close stellar encounters, that are most likely to happen early in the history of the solar system, when the sun was still in its birth cluster and the planets were still forming \citep{BRASSER-etal_2006}. This suggests that Sednoids are mainly constituted of big objects, which were rather unaffected by gas drag from the solar nebula \citep{BRASSER-etal_2007}. An early event is also supported by the fact that the Oort cloud would be severely (if not completely) depleted by such dramatic stellar passages. There is no such problem if the Oort cloud, created from the scattering of planetesimals at the time of the giant planet instability, was not even formed yet.
      \item[$\bullet$] The ``Oort cloud'' is the region where galactic tides and passing stars are the dominant orbital perturbations (possibly coupled with some planetary scattering during perihelion passages). The Oort cloud is thought to have been populated mostly by small bodies scattered away by the giant planets during their phase of instability. Galactic tides and passing stars then extracted the perihelion of scattered small bodies from the planetary region (see Sects.~\ref{sec:gt} and \ref{sec:stars}). As shown in Sect.~\ref{sec:inert}, the transition between planetary-dominated and galactic-dominated dynamics is fuzzy, located in the range $a\in[500,1600]$~au, and characterised by large-scale chaos. Beyond this limit, the Oort cloud is divided into its inner ($a\lesssim 20\,000$~au) and outer components ($a\gtrsim 20\,000$~au). In the outer Oort cloud, the orbital cycles described in Sect.~\ref{sec:gt} are so fast that the perihelion of small bodies can evolve from outside the scattering region to inside the orbit of Jupiter in less than one orbital period. As such, they are called ``jumpers'' by \cite{FOUCHARD-etal_2014}. In the inner Oort cloud, the orbital cycles are slow enough for bodies to make several successive perihelion passages in the planetary region if ever their trajectory leads them there. A few of them avoid catastrophic energy kicks and still manage to pass inside the orbit of Jupiter; they are called ``creepers'' by \cite{FOUCHARD-etal_2014}. However, the majority of inner-Oort-cloud bodies that get close to the planetary region are either purely ejected, or transferred to other regions: scattered disc, centaurs, or outer Oort cloud. In the latter case, their perihelion cycle is accelerated, making them safely pass inside the orbit of Jupiter \citep{KAIB-QUINN_2009}. Due to this mechanism, the inner and outer components of the Oort cloud equally contribute to the flux of observable long-period comets\footnote{Here again, the question of origin loses its meaning: most of the inner-Oort-cloud comets that become observable are actually first briefly transferred into the outer Oort cloud. Hence, the notion of origin largely depends on the time that we define as ``time zero''.}. They also contribute equally to the production of Halley-type comets, that become short-period due to repeated planetary kicks \citep{NESVORNY-etal_2017}. Among all Oort-cloud comets, only a fraction have trajectories that bring them near or into the planetary region under the effects of galactic tides (see Sect.~\ref{sec:gt}). The portion of the parameter space producing such trajectories is called the ``tidally active zone'' by \cite{FOUCHARD-etal_2011}. The tidally active zone would be completely depleted by now if it was not continuously refilled by the randomisation effects of passing stars (see Sect.~\ref{sec:stars}). Galactic tides and passing stars act therefore in synergy in the production of long-period and Halley-type comets \citep{RICKMAN-etal_2008,FOUCHARD-etal_2011}.
   \end{itemize}
   From this summary, it appears that the observed populations of trans-Neptunian objects, sculpted by the dynamical mechanisms described in this review article, are quite well understood today. And indeed, even if some questions remain open, much effort is now devoted to the precise quantitative aspects of the scenario, in particular its timing \citep{MORBIDELLI-NESVORNY_2019}.
   
   One of the open questions remaining concerns the origin of the significant orbital alignment of the most distant trans-Neptunian objects observed. A promising mechanism to explain such an alignment would be the shepherding effect of a distant ninth planet in the solar system (see the recent review by \citealp{BATYGIN-etal_2019}). However, this hypothetical planet has not been observed yet. The existence of ``Planet 9'' would not contradict the scenario outlined above, since it would have been created through the same mechanism as Sednoids. Planet 9 would not affect much the closest trans-Neptunian objects ($a\lesssim 70$~au), but it would dramatically modify the dynamical structure of the region located between about $200$ and $1500$~au in Fig.~\ref{fig:zones}. The inert zone would completely vanish, and wide pathways towards high perihelion distances would be opened in a complex web of mean-motion and secular resonances (see \citealp{BATYGIN-BROWN_2016,BEUST_2016,SAILLENFEST-etal_2017b,BATYGIN-MORBIDELLI_2017,HADDEN-etal_2018,LI-etal_2018}, and the review by \citealp{BATYGIN-etal_2019}). The notion of ``detached objects'' would lose its meaning, since all observed high-perihelion trans-Neptunian objects would have a very dynamic orbital evolution, which even includes inclination flips. Amid all this complexity, the orbits of observable scattered-disc objects with $a\gtrsim 250$~au would preferentially align with the orbit of Planet 9, as required to explain the current observational data. Importantly, the existence of Planet 9 would imply that planetary perturbations reach regions where the galactic tides are quite efficient, and that no weakly-perturbed transition regime exists. This has strong implications for the widely-used concept of ``original orbit'' that led \cite{OORT_1950} to predict the existence of the Oort cloud. The original orbit of a long-period comet is the orbit that it would have had at perihelion if there were no planets. It is estimated by propagating comets backward in time until they reach a distance where planetary perturbations can be neglected. If planets and external forces act together in a substantial portion of the trajectory, this concept should be redefined. A more complete description of the dynamics induced by ``Planet 9'' would be out of the scope of this review article. For more information about this hypothesis, we refer the reader to the articles cited above.
   
   Even when only taking into account the known planets of the solar system, the three-dimensional structure of the scattering region remains to be fully characterised, in particular at high inclinations. Moreover, we know that there is a threshold in semi-major axis above which resonances only appear as overlapping zones, probably because they are not strong enough to stand on their own. This limit is fixed to $a\approx 500$~au in Fig.~\ref{fig:zones}, but the precise position and the nature of this limit would deserve further investigations. It would also be worth introducing the variability of the galactic tide parameters into analytical models, in order to strengthen the numerical results obtained for instance by \cite{KAIB-etal_2011}. Finally, the complete stellar catalogue of \emph{Gaia} will soon allow us to build refined statistical models of stellar passages. Such models will yield much more precise estimates of the flux of bodies from the Oort cloud towards the scattered disc and the inner solar system. We will therefore obtain better constraints about the current structure of the Oort cloud, linked to its very formation process during the planetary instability \citep{FOUCHARD-etal_2018}.

\begin{acknowledgements}
   I thank Marc Fouchard for his support during the redaction of this review article. I am also very grateful to the two anonymous referees for their careful reading of the manuscript and their expert suggestions. This work was supported by the Programme National de Plan{\'e}tologie (PNP) of CNRS/INSU, co-funded by CNES.
\end{acknowledgements}

\bibliographystyle{aps-nameyear} 
\bibliography{reviewTNO}

\end{document}